\documentclass[10pt]{wlscirep}
\usepackage[utf8]{inputenc}
\usepackage[T1]{fontenc}

\title{Random walk diffusion simulations in semi-permeable layered media with varying diffusivity}
\author[1,2,*]{Ignasi~Alemany} 
\author[1,*]{Jan~N.~Rose} 
\author[1,4]{Jérôme~Garnier-Brun} 
\author[2,3,+]{Andrew~D.~Scott} 
\author[1,+]{Denis~J.~Doorly} 
\affil[1]{Department of Aeronautics, Imperial College London, South Kensington Campus, SW7 2AZ, London, UK}
\affil[2]{Cardiovascular Magnetic Resonance Unit, Royal Brompton Hospital, Sydney Street, SW3 6NP, London, UK}
\affil[3]{National Heart and Lung Institute, Imperial College London, SW3 6LY, London, UK}
\affil[4]{LadHyX, \'{E}cole Polytechnique, Paris, France}
\affil[+]{Contributed equally as joint senior authors}
\affil[*]{Authors contributed equally to this work: ignasi.alemany18@imperial.ac.uk and jan.rose14@imperial.ac.uk}

\keywords{Diffusion, Key, word}

\usepackage[utf8]{inputenc}

\usepackage{xcolor}
\usepackage{soul}

\usepackage{placeins} 

\usepackage{graphicx}
\graphicspath{{floats/}} 
\setkeys{Gin}{width=\textwidth} 
\makeatletter
\renewcommand*{\fps@figure}{t}
\renewcommand*{\fps@table}{b}
\makeatother

\usepackage{tabularx}
\newcolumntype{C}{>{\centering\arraybackslash}X}

\usepackage{algorithm}
\usepackage{algpseudocode} 

\usepackage{enumitem}

\usepackage{amssymb}
\usepackage{latexsym}
\usepackage{amsmath}
\usepackage[centercolon]{mathtools}
\usepackage{comment}

\usepackage{diffcoeff}
\diffdef{}{
op-symbol = \mathrm{d}
}
\diffdef{p}{
left-delim = \left. ,
right-delim = \right| ,
subscr-nudge = 0mu ,
}
\newcommand{\del}[1]{\ensuremath{\dl.delta.{#1}}} 

\usepackage{xfrac}

\renewcommand{\matrix}[1]{\boldsymbol{#1}}
\renewcommand{\vector}[1]{\boldsymbol{#1}}
\newcommand{\gradient}[1]{\vector{\nabla}{#1}}
\newcommand{\divergence}{\vector{\nabla}\cdot}

\DeclarePairedDelimiter{\absolute}{\lvert}{\rvert}
\DeclarePairedDelimiterXPP{\order}[1]{\mathcal{O}}{(}{)}{}{#1}
\DeclarePairedDelimiterXPP{\defint}[3]{}{[}{]}{_{#2}^{#3}}{#1} 

\usepackage[binary-units=true]{siunitx}
\DeclareSIUnit\unitL{\micro\meter}
\DeclareSIUnit\unitT{\milli\second}
\DeclareSIUnit\unitK{\unitL\per\unitT}
\DeclareSIUnit\unitD{\unitL\squared\per\unitT}
\DeclareSIUnit\unitB{\unitT\per\unitL\squared}

\usepackage{hyperref}
\usepackage[noabbrev]{cleveref} 

\usepackage{comment}
\usepackage[mathlines,right]{lineno} 
\usepackage{setspace} 

\usepackage{etoolbox} 

\newcommand*\linenomathpatch[1]{%
  \expandafter\pretocmd\csname #1\endcsname {\linenomath}{}{}%
  \expandafter\pretocmd\csname #1*\endcsname{\linenomath}{}{}%
  \expandafter\apptocmd\csname end#1\endcsname {\endlinenomath}{}{}%
  \expandafter\apptocmd\csname end#1*\endcsname{\endlinenomath}{}{}%
}
\newcommand*\linenomathpatchAMS[1]{%
  \expandafter\pretocmd\csname #1\endcsname {\linenomathAMS}{}{}%
  \expandafter\pretocmd\csname #1*\endcsname{\linenomathAMS}{}{}%
  \expandafter\apptocmd\csname end#1\endcsname {\endlinenomath}{}{}%
  \expandafter\apptocmd\csname end#1*\endcsname{\endlinenomath}{}{}%
}
\expandafter\ifx\linenomath\linenomathWithnumbers
  \let\linenomathAMS\linenomathWithnumbers
  \patchcmd\linenomathAMS{\advance\postdisplaypenalty\linenopenalty}{}{}{}
\else
  \let\linenomathAMS\linenomathNonumbers
\fi
\linenomathpatch{equation} 
\linenomathpatchAMS{gather}
\linenomathpatchAMS{multline}
\linenomathpatchAMS{align}
\linenomathpatchAMS{alignat}
\linenomathpatchAMS{flalign}

\usepackage[title]{appendix}
\AtBeginEnvironment{appendices}{%
\appendix 
}

\begin{abstract}
In this paper we present analytical and random walk based solutions to diffusion in semi-permeable layered media with varying diffusivity. We propose a new random walk transit model (hybrid model) based on treating the membrane permeability and the change in diffusion as two infinitesimal separate phenomena.  By conducting an extensive analytical flux analysis, the performance of our hybrid model is compared with a commonly used membrane model (reference model). We numerically demonstrate the limitations of the reference model and show the capability of our new model to overcome these restrictions. The suitability of both random walk transit models for the application to simulations of the diffusion tensor cardiovascular magnetic resonance (DT-CMR) is assessed in a histology-based domain. We consider a larger range of permeabilities to show the potential of our model to other possible applications beyond biological tissue.
    
\end{abstract}

\begin{document}

\flushbottom
\maketitle

\section{Introduction}

The understanding of the diffusion of fluid particles within a material consisting of multiple compartments separated by semi-permeable barriers or membranes is important in a vast number of areas such as heat transfer problems~\cite{Hickson2009,Vignoles2016}, mathematical modelling in finance~\cite{Decamps2004} or social dynamics~\cite{Berestyki2016}, astrophysics~\cite{Zhang2000}, the study of porous media~\cite{Berkowitz2006,DiffusionNanoporous}, and diffusion-weighted imaging~(DWI). DWI is a unique magnetic resonance imaging technique that provides measures relating to the average microscopic structure within a macroscopic imaging voxel by measuring the displacement of water molecules due to self-diffusion over a given time~(the diffusion time~$\Delta$)~\cite{Stejskal1965,LeBihan2003}. DWI-based methods are widely used in neuroscience for determining white matter pathways, for example via the primary eigenvector of the $3 \times 3$~diffusion tensor~\cite{Basser1994,LeBihan2012} which aligns with the long axis of the neurons. More recently, DWI techniques have gained popularity for cardiac imaging, where they can be used to investigate the unique variations in the arrangement of heart muscle cells in space and in time as the heart contracts~\cite{NiellesVallespin2020}.

Monte Carlo simulations are a well-established method for investigating the relationship between the properties of the microscopic structure of biological tissues and the apparent diffusivity which would be measured in DWI methods~\cite{Fieremans2018}. These computational simulations are becoming increasingly more realistic in terms of geometric fidelity~\cite{Yeh2013,Palombo2019,Rose2019}. Compartment models, which assume the tissue consists of a number or distribution of compartment sizes and consider the DWI signal contribution of each compartment separately, have reached a point of maturity, but are limited to tissues with no/low membrane permeability or short diffusion times. In cardiac tissue, however, diffusion times are of the order of~\SI{1}{\second} when employing the commonly used Stimulated Echo Acquisition Mode~\cite{Reese1995} technique due the synchronisation with the cardiac period. As a result, the mean distance displaced by a molecule during an experiment covers multiple compartment lengths and the membranes of the typically well-mixed myocardial cells~\cite{Bruvold2007} may no longer be considered impermeable.

Exchange of walkers through barriers is commonly modelled via a transit probability, where an attempt to cross the barrier is either rejected or permitted randomly based on a threshold probability. The value of this is dependent on some (or all) of the tissue properties and its choice aims to ensure the membrane permeability is accurately represented. Powles et al.~\cite{Powles1992}~derived a formula for the transit probability of walkers on a lattice with constant step size. Szafer et al.~\cite{Szafer1995}~considered a grid of 3D rectangular cells on a regular grid, allowing for different intra- and extra-cellular diffusivities. A recent model \cite{Fieremans2010,Lee2020} (reference model) attempts to improve and extend the performance of a previous published transit model \cite{Powles1992}. The latter requires a strict limit on the maximum time step permitted in the random walk, which may pose numerical challenges when long diffusion times are considered or a large parameter space is to be investigated.

A number of analytical approaches to the problem of diffusion also exist. Data fitting models attempt to compose the observed DWI signal as a linear combination of analytical shapes like spheres or ellipsoids for which the contribution is known~\cite{Panagiotaki2012}. This allows for the inference of cell sizes from the measured data. Originally developed for impermeable membranes, this approach was extended by~Karger e.al~\cite{Karger1985} to account for exchange between compartments. While these models offer an easy way to explain macroscopic observations through integral quantities like the diffusion signal, they do not allow for deeper insights into the diffusion processes themselves. By reducing the problem to 1D, analytical solutions for the diffusion propagator can be found. This was first suggested by Tanner et al.~\cite{Tanner1978} to estimate the DWI signal in a system of equi-spaced parallel plates. Very recently, Moutal et al.~\cite{Moutal2019} presented a semi-analytical method to obtain the particle density distribution anywhere in a domain with arbitrary barriers. 

In this work, we study 1D~diffusion through semi-permeable membranes and show the numerical limitations of the previous mentioned reference transit model \cite{Fieremans2010,Lee2020}. We propose a new model (hybrid model) based on treating the membrane interface and the diffusivity discontinuity as two separate probabilities. We analyse the behaviour of both transit model by comparing the fluxes through the membrane with a semi-analytical solution. A parameter study reveals the errors in the numerical results and demonstrates that the hybrid model limitations are numerically far less restrictive. Finally, we quantify the impact of our findings by calculating the difference in DWI signal obtained using both transit model on a realistic histology-based domain with a wide range of permeability values. This allows us to assess and compare the suitability of both models on diffusion tensor cardiovascular magnetic resonance (DT-CMR) and to other potential applications.  


\section{Methods}

We describe a method to find (semi-)analytical solutions of the diffusion equation
\begin{equation}
    \diff.p.{U(\vector{x}, t)}{t} = \divergence\Big[ D(\vector{x}) \gradient{U(\vector{x}, t)} \Big]
    \label{eqn:diffusion}
\end{equation}
in 1D~layered media, i.e.~$\dot{U} = \diff.p.*{[D \diff.p.{U}/{x}]}{x}$ with piece-wise constant diffusion coefficient~$D(x)$.

In this work, we consider domains as arrays of compartments with constant properties. \Cref{fig:rw_stepping} provides an example of one such arbitrary domain and explains the nomenclature. The \namecref{fig:rw_stepping} also shows cardiomyocytes obtained via confocal microscopy imaging, typically employed to get detailed structural information about the tissue microstructure. Image data like this often informs (computational)~models of diffusion in biological tissue and aids the inference of tissue properties required from~DWI. Considering that, for typical diffusion length scales, diffusion in the vertical/longitudinal direction~(parallel to the cardiomyocytes) is much less restricted than in the perpendicular direction, the problem may be reduced to~2D for many applications. Even 1D~simulations provide new insights, especially in studying the effect of permeability.


All parameters uniquely defining the problem (compartment barrier locations~$b_k$ and corresponding sizes~$L_k$, diffusivities~$D_k$, barrier permeabilities~$\kappa_k$) can be arbitrarily chosen, which allows us to consider histology based domains. The choice ~$D_k$ permits the modelling of different compartments as intra-cellular~(ICS) or extra-cellular space~(ECS, which may be interstitial or intravascular). Even though, at the molecular level, walkers experience the same self-diffusion\footnote{The diffusion coefficient of a given fluid is only dependent on the temperature.}, a mesoscopic model of a reduced bulk diffusivity accounts for apparent hindrance due to intra-compartmental tissue characteristics. Interfaces like membranes or intercalated discs may either be modelled with reduced~$D$ and small~$L$, or as barriers with appropriate permeability~$\kappa$. We set a zero-flux Neumann boundary condition~($\gradient{U} = 0$) for this investigation as walkers do not vanish in biological tissue. 

\subsection{Random walk transit models for permeable membranes}

The solution to the diffusion equation~\labelcref{eqn:diffusion} may be obtained by considering an ensemble of massless, non-interacting particles performing a random walk. $U(x)$~ denotes the probability of a particle being located at a given position. Below, we describe the random walk process for a single particle/walker, which is repeated $N_p$~times per experiment. 

\subsubsection{A single random walker}

We consider a walker with subscript~$p$. At each time step~$\del{t}$, its position~$x_p$ is updated through a series of sub-steps~$\del{x}_n$ that depend on the local environment:
\begin{equation}
    x_p(t+\del{t}) = x_p(t) + \sum_{n}{ \del{x}_n } \, .
\end{equation}
At the beginning of a time step, a random step vector~$\del{x}$ is drawn with equal probability of moving in direction~$+x$ or~$-x$. For diffusion away from barriers, a single step~$\del{x} = \pm \sqrt{2 D \del{t}}$ is performed. Interaction with barriers introduces additional sub-steps, however in this work we consider a maximum of a single barrier crossing per time step. This imposes a limit on the possible time steps, namely $\del{t} < \min_k\left(\frac{L_k}{2 D_k}\right)$.

\begin{figure}
    \centering
    \includegraphics{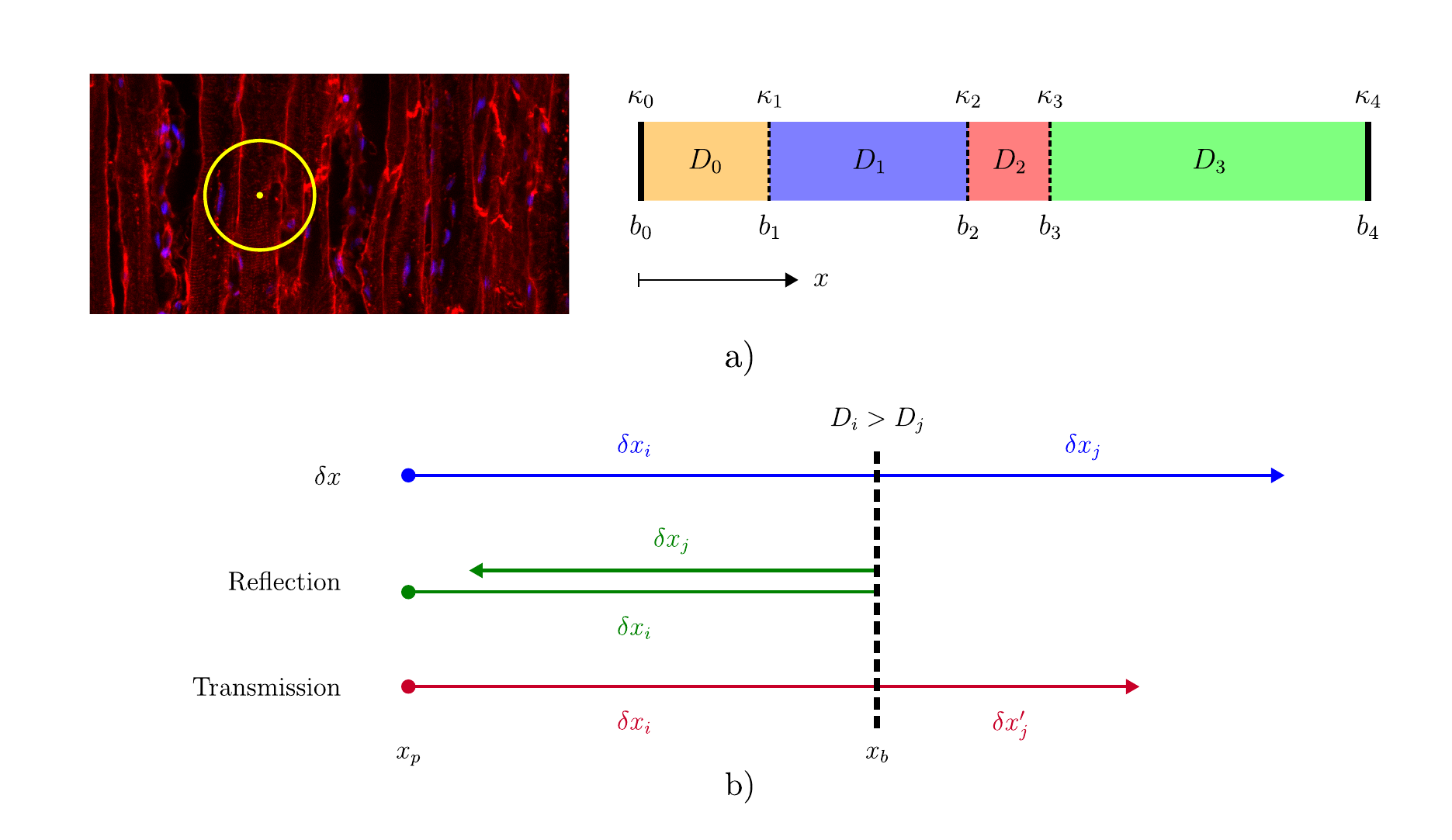}
    \caption{\textbf{a}) \textit{Left}: Confocal fluorescence microscopy image of cardiomyocytes running vertically. The mean diffusion distance~$\dl.Delta.x \approx \SI{30}{\unitL}$ over \SI{100}{\unitT} is indicated as the radius of the yellow circle. \textit{Right}: Schematic of an example domain with $m = 4$~compartments with indices~$k$. Each compartment has two barriers with their corresponding locations ~$b_{k}$, diffusivites~$D_{k}$, and permeabilities~$\kappa_{j}$. Note that the domain ends enforce the zero-flux boundary conditions~($\kappa_{0} = \kappa_{m} = 0$).
    \textbf{b}) Illustration of the behaviour of a single walker at~$x_p$ performing a random step~$\del{x}$ towards a barrier at~$x_b$. Initially, the step is divided into $\del{x}_i$ and $\del{x}_j$. Depending on the transit decision, the walker is either reflected elastically~($x = x_p + \del{x}_i - \del{x}_j$) or enters the new compartment with~$D_j < D_i$. In the latter case, the remaining step after transit is modified to~$\del{x}'_j$ following \cref{eqn:afterStep} such that entering a compartment with lower/higher diffusivity decreases/increases~$\del{x}_j$.}
    \label{fig:rw_stepping}
\end{figure}

If a walker attempts to cross from compartment~$i$ to compartment~$j$ through a barrier at $x_b$, the step is divided into~$\del{x} = \del{x}_i + \del{x}_j$ such that~$\del{x}_i = x_b - x_p$. This is illustrated in \cref{fig:rw_stepping}. The interaction is resolved by first computing a probability of transit~$p_t$ and then drawing a random number~$\mathcal{U} \in [0,1)$ to compare to~$p_t$. Transit occurs if~$\mathcal{U} < p_t$, otherwise the walker is reflected elastically. Upon entering a new compartment with different diffusivity~$D_j \neq D_i$, the remaining step length~$\del{x}_j$ over the duration~$\del{t}_j = \frac{\del{x}_j}{\del{x}}\del{t}$ needs to be adjusted to preserve a constant net step size~\cite{Szafer1995}:
\begin{equation}
    \del{x}'_j = \del{x}_j \sqrt{\frac{D_j}{D_i}}
    \label{eqn:afterStep}
\end{equation}

In addition to the above step modification, another condition needs to be satisfied. The probabilities of crossing from either side to the other, i.e.~$p_{t, (i \to j)}$ and~$p_{t, (j \to i)}$, must be related by
\begin{equation}
    \frac{p_{t, (i \to j)}}{p_{t, (j \to i)}} = \sqrt{\frac{D_j}{D_i}} \, .
    \label{eqn:interface_reflection}
\end{equation}
This ``interface reflection'' is required even as~$\kappa \to \infty$\cite{Szafer1995}.

\subsubsection{Hybrid model}
\label{sec:hybrid_model}

A number of "transit models" have been proposed to calculate the probability of transit~$p_t$ based on properties of the membrane and tissue. The ultimate aim of any such model is to accurately represent the \emph{leather} boundary condition~\cref{eqn:leather_BC} \cite{Powles1992,Tanner1978}.
\begin{equation}
    D_i \diffp{U}{x}[L] = \kappa_{i,j} \left( \left. U \right|_R - \left. U \right|_L \right)
    \label{eqn:leather_BC}
\end{equation}
Here, $U$~is the particle density and the evaluation limits~$L$ and~$R$ indicate that the concentration and its gradient should be evaluated infinitesimally to the left (on the side of compartment~$i$) or right (compartment~$j$) of the interface. Fick's first law~\cite{Fick1855} states that the flux~$\vector{J}$ is related to the gradient in concentration by~$\vector{J} = - D \gradient{U}$, where the sign indicates the direction of the flux vector with magnitude~$J$. Since there is no accumulation of concentration/walkers inside the barrier, the flux must remain the same on either side, i.e.
\begin{equation}
    D_i \diffp{U}{x}[L] = D_j \diffp{U}{x}[R] \, .
    \label{eqn:flux_BC}
\end{equation}

The case of a finite membrane permeability with constant diffusivity values  (i.e.~$D_i = D_j$) is well-studied on a discrete lattice of equidistant points \cite{Powles1992}. A recent transit model\cite{Fieremans2010,Lee2020} (reference model) extended this approach to particles located at arbitrary positions~$x_p$ in the vicinity of the membrane. By neglecting higher-order terms in the derivation and still considering constant diffusivity (compare \cref{sec:ModelDerivation}), it is possible to derive a transit probability $p_t$ that works for different diffusivites provided that the time step is small enough such that ~$p_t \ll 1 < 0.01$ \cite{Fieremans2010}.
\begin{equation}
    p_{t, (i \to j)} = \frac{2 \kappa_{i,j} \del{x}_i}{D_i + 2 \kappa_{i,j} \del{x}_i} \, ,
    \label{eqn:pt_Fieremans}
\end{equation}
where $\del{x}_i$~is the distance from~$x_p$ to the barrier~$x_b$. With the motivation of lifting this step restriction, we propose a new transit model (hybrid model) based on treating the diffusivity gradient and the membrane permeability as two independent factors. This new transit model considers an infinitesimal space between the membrane and the diffusivity change such that the overall probability can be computed as the product of these two.
\begin{equation}
    p_{t, (i \to j)} = p_{m, (i \to j)} \cdot p_{d, (i \to j)}
    \label{eqn:pt_hybridmodel}
\end{equation}

Where $p_{\text{m}}$ and $p_{\text{d}}$ are the membrane probability considering constant diffusivity and the probability of a particle when transitioning between two different media. The probability $p_{\text{d}}$ is presented as an elegant interpretation of the behaviour of random walkers when transitioning between media of different viscosities. \cite{Maruyama2017}.
\begin{equation}
    p_{d, (i \to j)} = \min \left(1, \sqrt{\frac{D_j}{D_i}} \right) \, ,
    \label{eqn:pt_Maruyama}
\end{equation}

\begin{figure}  
    \centering
    \includegraphics{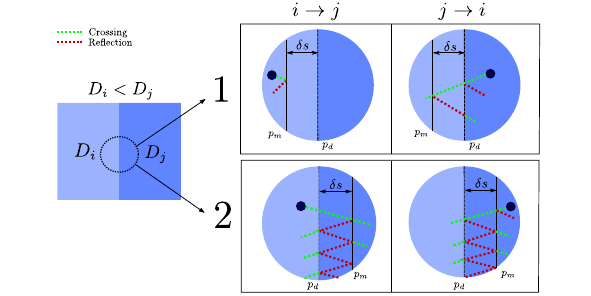}
    \caption{Illustration of the two different configurations depending on where the interface membrane ($p_m$) is placed. Configurations 1 and 2 place the membrane in the "slow side" (low diffusivity region) and "fast side" (high diffusivity region) respectively.}
    \label{fig:hybrid_model}
\end{figure}

Substituting $p_{\text{m}}$ and $p_{\text{d}}$ with \cref{eqn:pt_Maruyama} and \cref{eqn:pt_Fieremans} we obtain the overall probability of the new transit model. As observed in \cref{fig:hybrid_model}, this hybrid model presents two possible configurations depending on the location of $p_m$. The first and second option place the membrane in the low and high diffusivity region respectively. We observe that positioning $p_m$ in the "fast" side (high diffusivity) leads to infinite reflections in between the two membranes ($\delta{s}$). The model considers $\delta{s}\to 0$ because considering a small $\delta{s}$ would set a high step size restriction. Due to the fact that there is no physical compartment and infinite reflections when placing $p_m$ in the "fast side", the model is only valid when the membrane $p_{m}$ is placed in the "slow side" (low diffusivity region). Even though locating $p_m$ in the fast side is not feasible when $\delta{s}\to 0$, in \cref{S_Appendix_hybridmodel}, we mathematically show and validate that the infinite reflections that occur when placing the $p_m$ in the high diffusivity region ("fast side") lead to the same overall probability as placing the membrane $p_m$ in the slow slide.

\subsection{Flux analysis}

We investigate the transient and steady state behaviour of~$U(x, t; x_0)$ in a very simple representative domain using the reference and new hybrid transit models mentioned in \cref{sec:hybrid_model}. We compare the fluxes through the permeable barrier to an analytical solution. Analysing the fluxes offers more insights into the exchange process, which is modelled as a random process, than looking at concentrations on either side of the interface. The flux as an integral quantity can be compared easily with the analytical solution and does not rely on histogram binning, which is prone to noise.

\subsubsection{Simple domain}
\label{sec:simple_domain}

Very few estimates exist in literature for the free diffusivity in the myocardium and/or the permeability of the  cardiomyocyte membrane. A recent study \cite{Seland2007} observed diffusivity values of $\SI{1.2}{\unitD}$ and $\SI{3}{\unitD}$ in the intra- and extra-cellular space of heart rat cells. A recent numerical study~\cite{Rose2019} performed simulations of a histology-based domain for a range of diffusivities and found apparent diffusion coefficients~(ADC) similar to what is commonly observed in-vivo. The values for~$D_\text{ECS}$ had a range of~\SIrange{1}{3}{\unitD}, with~$D_\text{ICS}$ varying between~\SI{1}{\unitD} and~$D_\text{ECS}$. In general it should be expected that the compartment-specific bulk diffusivity are lower than the free diffusivity of water and $D_\text{ICS}<D_\text{ECS}$ due to the concentration of sub-cellular structures.The cell membrane permeability may be estimated from measurements of apparent exchange rate ($1/\tau_\text{ex}$). The exchange time constant~$\tau_\text{ex}$~ is related to the cell membrane permeability via~$\tau_\text{ex} = \alpha\;(V/S)/\kappa$, where $\alpha$~is the extra-cellular volume fraction~(ECV) and ~$V$ and~$S$ the volume and surface area of the cell respectively. Exchange rates are commonly reported in a range of~\SIrange{6}{30}{\Hz}~\cite{Seland2007,CoelhoFilho2013}, but have been found as high as~\SI{50}{\Hz} for healthy leg muscles of rats~\cite{Sobol1990}.

We consider a domain with a single permeable barrier separating two compartments of different diffusivities,~$D_L$ and~$D_R$, representing intra- and extra-cellular space respectively. This allows us to isolate a single interface and avoid confounding effects from other compartments. The parameter choice aims to be representative of cardiac tissue. We use~$D_\text{L} = \SI{0.5}{\unitD}$ and~$D_\text{R} = \SI{2.5}{\unitD}$ as compartment diffusivities expected in intra- and extra-cellular space. The membrane permeability is set to $\kappa = \SI{0.05}{\unitK}$ based on an estimated $\tau_\text{ex}^{-1} = \SI{30}{\Hz}$. The compartment lengths are considered to be equal in both sides $L=\SI{20}{\unitL}$

\subsubsection{Analytical Solution}
\label{sec:ana}

In order to compare the performance of both transit models we have the necessity to obtain an analytical solution. A recent study \cite{Moutal2019} presented an elegant transcendental equation~$F(\lambda) \coloneqq 0$ that is to be solved in order to obtain the eigenvalues~$\lambda$ of the diffusion operator. The eigenvalues are the roots of~$F(\lambda^\star)$ where $\lambda^\star$~is an auxiliary variable considered when $F$~is evaluated on a continuous range. The solution assumes the time and space variables in \cref{eqn:diffusion} are separable, resulting in the ansatz
\begin{equation}
    U(x, t) = u(x) e^{-\lambda t} \, .
\end{equation}
For piecewise-constant~$D_k$, the eigenvalue problem is reduced to a Helmholtz equation
\begin{equation}
    D_k \, u'' + \lambda u = 0 \quad \forall x \in [b_k, b_{k+1}] \, .
    \label{eqn:Helmoltz_Di}
\end{equation}
Using the general solution to this ODE, which is a function of the eigenvalues, one can construct a transcendental equation ensuring that all boundary conditions are satisfied simultaneously. This is achieved through a series of matrix multiplications that link compartments together~\cite{Moutal2019}. The problem of finding eigenvalues is therefore reduced to that of finding the roots of that transcendental equation
\begin{equation}
\label{eqn:F(lambda)}
    F(\lambda)
    \coloneqq
    \begin{bmatrix}
        \kappa_m/\sqrt{\lambda D_{m-1}} & 1\\
    \end{bmatrix}
    %
    \matrix{R}_{m-1}(\lambda)
    \left(
    \prod_{k=0}^{m-2}{
        \matrix{M}_{k,k+1}(\lambda)
        \matrix{R}_k(\lambda)
    }
    \right)
    \begin{bmatrix}
        1 \\ \kappa_0/\sqrt{\lambda D_0}
    \end{bmatrix}
    = 0
\end{equation}
with auxiliary matrices
\begin{subequations}
\begin{equation}
    \matrix{R}_k(\lambda) =
    \begin{bmatrix}
    \cos( L_k \sqrt{\lambda/D_k} ) & \sin( L_k \sqrt{\lambda/D_k} ) \\
    -\sin( L_k \sqrt{\lambda/D_k} ) & \cos( L_k \sqrt{\lambda/D_k} )
    \end{bmatrix}
    \, ,
\end{equation}
\begin{equation}
    \matrix{M}_{k,k+1}(\lambda) = 
    \begin{bmatrix}
        1 & \sqrt{\lambda D_k}/\kappa_{k,k+1} \\
        0 & \sqrt{D_k/D_{k+1}}
    \end{bmatrix}
    \, .
\end{equation}
\end{subequations}

For a sensible choice of permeabilities for internal barriers~($0 < \kappa_k < \infty$, i.e.~no trivial cases), there exist a countably infinite number of real eigenvalues, all of which are non-negative, ordered, and simple~\cite{Moutal2019}:
\begin{equation}
    0 \leq \lambda_{1} < \lambda_{2} < \, \dots \, , \quad \lambda_{n} \to \infty
\end{equation}
The eigenvalues of higher order have diminishing importance with increasing solution time~$t$. For the domains in this work, we use a truncation point of the order of ~$n \leq \order{\num{1e3}}$. The solution is then computed by evaluating and summing the eigenmodes~$\nu_i(x)$ throughout the domain at several linearly-spaced points~$x_q \in [0, b_m]$.    
\begin{equation}
    U(x_q, t) \approx \sum_{n=1}^{N}{ e^{-\lambda_n t} \nu_n(x_q) \nu_n(x_{q,0}) }
    \label{eqn:ana_series}
\end{equation}
The initial condition is always given by a delta Dirac function:
\begin{equation}
    U(x_q, 0) = \delta(x_q - x_{q,0})
    \label{eqn:ana_initialcondition}
\end{equation}
As described above, the eigenvalues (roots of \cref{eqn:F(lambda)}) are found numerically up to~$\lambda_\text{max}$. The solution~$U(x_q, t)$ is, therefore, semi-analytical. As a consequence, numerical errors introduced by the root finding procedure manifest themselves as errors in the solution. In our simple and complex domains, $\lambda_{max}=1000$ has been observed to be enough for an accurate and smooth solution. Due to the linearity of the diffusion equation, a uniform initial condition in a certain domain region can be solved by adding and normalising all the solutions obtained by several delta Dirac functions within the interested interval.

\subsubsection{Transient and steady state analysis}
\label{methods_transient_steady_simple}
We analyse the flux of particles crossing the membranes that is represented from the flux boundary condition in  \Cref{eqn:leather_BC}. This equation relates the flux~$J$ with the permeability and the difference between concentrations across the barrier. The units of~$J$ are concentration (fraction of walkers) per unit time and area, but we omit the latter such that~$[J] = \si{\per\unitT}$. This flux boundary condition, allows us to compute the instantaneous analytical flux evaluating U on either side of the discontinuity and the instantaneous numerical flux by measuring the fraction of particles that cross the barrier for the each specific time step. The net instantaneous flux is obtained each time step  by subtracting the fluxes from either side (left/right) and the cumulative flux (flux integral) throughout the simulation.

Based on the analytical and numerical flux comparison, we perform two different types of analysis to evaluate the efficacy of the transit models. We implement a steady and transient state analysis by considering different initial conditions. A uniform concentration throughout the domain is considered for the steady-state and a partial uniform distribution for the transient state. We determine the relative error $\epsilon_{global}$ for both transit models using the cumulative flux ~$\mathcal{J}(t) = \int_{0}^{t}{J(\tau) \dl2{\tau}}$ which represents the net concentration of walkers that has crossed the membrane up until $t$ and approaches $0.5$ as $t \to \infty$ to match the steady state solution. This global relative error $\epsilon_{global}$ measures the area in between the analytical and numerical solution during the entire simulation.We utilise this global relative error to investigate the time step dependence and compare the performances of our new hybrid model and the reference model \cite{Fieremans2010} under the influence of different permeability and diffusivity values.

\subsection{Diffusion Weighted Imaging (DWI)}

\subsubsection{Synthesised histology-based domain}
\label{sec:methods_synthesised_domain}
The domain has been obtained from sections of swine myocardium, cut perpendicular to the long axis of the cardiomyocytes. We have used automatic segmentation developed in previous work\cite{ISMRM2018:Rose} to obtain a distribution of cell sizes. We have utilised this previous work to find statistics parameters (mean cross-sectional area, standard deviation) to recreate a synthesised histology-based domain assuming a circular cross-sections for cardiomyocytes \cite{Tracy2011}. The cell areas are converted to diameters which we use as intra-cellular compartment lenghts in the 1D domain. The extracellular space is recreated considering a uniform distribution.

\subsubsection{Random walk simulations in the histology-based domain}
\label{methods:histology_domain}
For the histology domain simulations, we use $N_p = \num{1e6}$~walkers and a time step of~$\del{t} = \SI{1.5}{\unitT}$ as it has been observed to be enough to converge to an accurate solution. These parameters intentionally exceed the small time step required for accurate handling of transit using the model in \cref{eqn:pt_Fieremans}, while still limiting walkers to a single barrier interaction per step (since $\min(L_\text{ECS}) \geq \SI{2}{\unitL}$ and $D_\text{ECS}=\SI{0.5}{\unitD}$). Random walk simulations are run for~$t = \SI{1000}{\unitT}$ and the analytical solution is evaluated for the same space and time parameters. We use the semi-analytical solution described in \cref{sec:ana} to obtain \num{1000} eigenvalues with $\lambda^\star \in [0, 500]$. Conversely to \cref{sec:rw_transient}, the transient solution~$U(x, t; x_0)$ is obtained by seeding all the walkers at the centre of the domain, while the steady-state solution is initialised by seeding the walkers uniformly. We use~$D_\text{ICS} = \SI{0.5}{\unitD}$ and~$D_\text{ECS} = \SI{2.0}{\unitD}$ and we set the permeability to $\kappa = \SI{0.05}{\unitK}$ for all cell membranes.

\subsubsection{DWI and other potential applications}
\label{methods:application_dwi}
One important application of Monte Carlo random walk simulations in complex domains is in understanding diffusion weighted imaging (DWI). We perform numerical simulations of DWI in the histology-based domain using the reference model \cite{Fieremans2010,Lee2020} and our new transit model proposed in \cref{sec:hybrid_model}. In DWI, radio-frequency pulses are used to excite hydrogen spins and magnetic field gradients change their precession according to their gyromagnetic ratio~$\gamma = \SI{267.5e6}{\radian\per\second\per\tesla}$. Each spin/walker~$p$ collects precessional phase
\begin{equation}
    \phi_p(t) = \gamma \int_{0}^{t}{ G(\tau) x_p(\tau) \dl2{\tau} }
\end{equation}
during the random walk, subject to two symmetric gradients of strength~$G$ and duration~$\delta$. At readout/time of echo, the resulting signal~$S$ is the Fourier transform of the medium average diffusion propagator and the signal magnitude is thus related to the distances that the spins have diffused during time~$\Delta$.

The analytical and random walk methods presented previously allow us to solve for the diffusion propagator. By means of the narrow pulse approximation~(NPA)~\cite{DiffusionMRI:Callaghan,Grebenkov2014} it is possible to estimate the DWI signal directly. This assumes that the gradients are applied instantaneously, i.e.\ $\delta \to 0$ as the wave number $q(t) = \gamma \int_{0}^{t}{ G(\tau) \dl2{\tau} }$ remains finite ($=\gamma G \delta$). In DWI, the strength and timing of the gradients is reflected through a parameter called b-value. In the case of NPA, the b-value is calculated as $b=q^2 \Delta$

We seed the walkers uniformly in the histology-based domain and we let them diffuse for a period of 1 second. We analyse and compare the signal attenuation and apparent diffusion coefficient (ADC) with analytical results for a large time step of $dt = \SI{1.5}{\milli\second}$ and a small time step $dt = \SI{0.01}{\milli\second}$. In order to provide insights in other potential applications, we consider a wide range of permeabilities ($0$ to $\SI{1}{\unitK}$) apart from the biological ambit. We have considered a b-value of $\SI{1}{\milli\second\per\micro\meter\squared}$ for all the simulations. Due to the narrow pulse approximation (NPA), the random walk and analytical signal ($S_\text{rw}, S_\text{ana}$) can be directly computed using the following equations:
\begin{equation}
    S_\text{rw}(\Delta, q) = \frac{1}{N_p} \absolute*{ \sum_{p}^{N_p}{ e^{-\imath \twomu q \twomu \left( x_p(\Delta)-x_p(0) \right)} } }
    \label{eqn:NPA_mcrw}
\end{equation}
\begin{equation}
    S_\text{ana}(\Delta, q) = \frac{1}{\sum_{k}{L_k}} \sum_{n}{ e^{-\lambda_n \Delta} \absolute*{ \int_{\Omega}{ \nu_n(x) e^{\imath q x} \dl2{x} } }^2 }
    \label{eqn:NPA_ana}
\end{equation}

Here, $\imath$~denotes the imaginary unit~$\sqrt{-1}$. The analytical diffusion signal is calculated using an expression derived for uniform initial seeding~\cite{Moutal2019}. In \cref{eqn:NPA_ana}, we approximate the integral numerically using trapezoidal integration over the finely-discretised domain. Finally, the apparent diffusion coefficient~($\mathrm{ADC}$) is derived from the signal attenuation through $\mathrm{ADC} = -\ln(S/S_0)/b$ (assuming $S_0 = 1$).

\section{Results}
\label{sec:rw}

\subsection{Analysis of the steady-state}
\label{sec:rw_steadystate}
\begin{figure}
   \centering
    \includegraphics{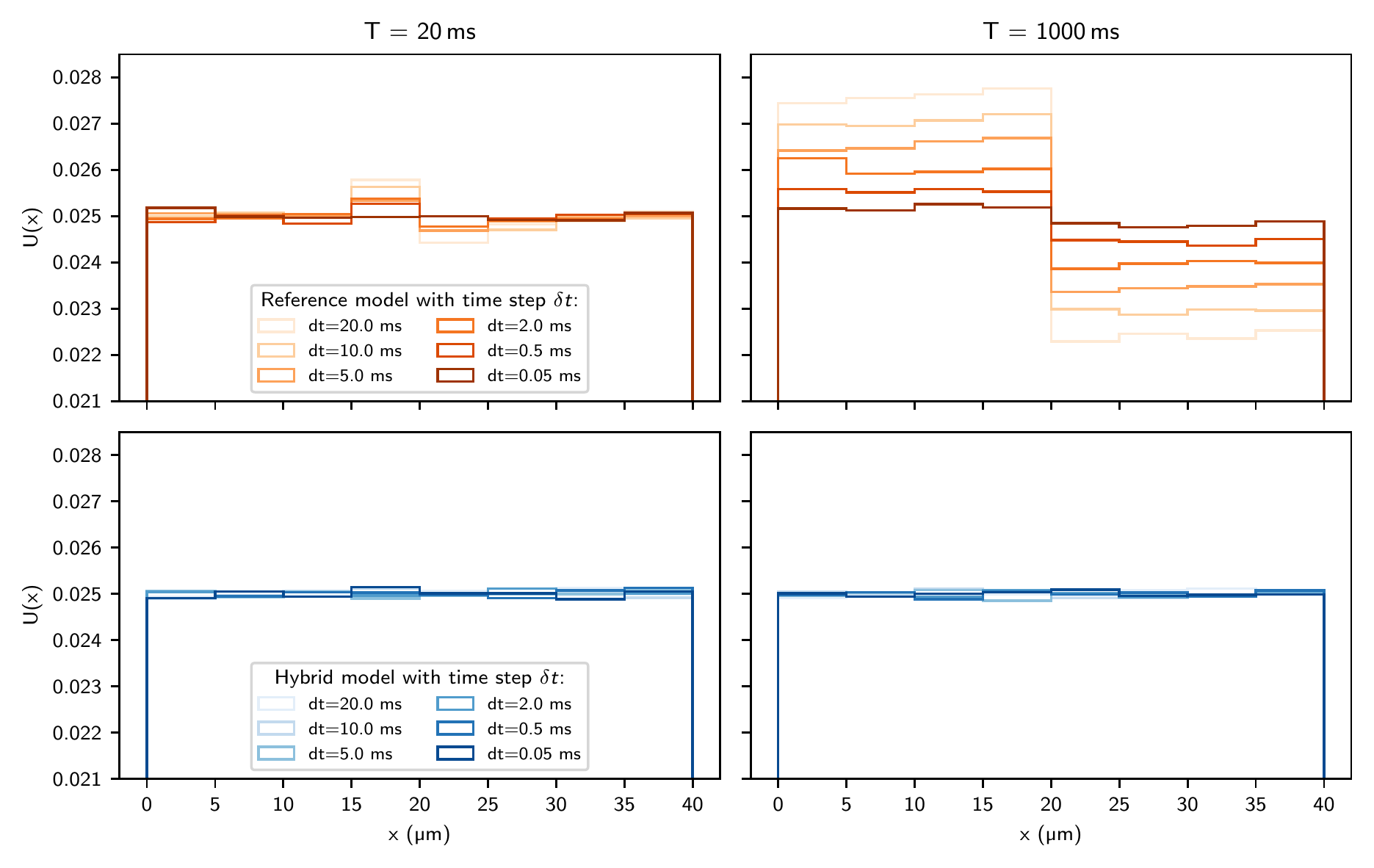}
   \caption{Histograms of walker positions after random walk simulations of the steady state. Initial positions were sampled from a uniform distribution to seed walkers with a constant density throughout the domain. We applied the reference membrane model (Top) described in \cref{eqn:pt_Fieremans}, with permeability~$\kappa = \SI{0.05}{\unitK}$, and the new hybrid model (Bottom) from \cref{eqn:pt_hybridmodel}. Simulations were performed with varying step sizes~$\del{t}$ for a short and a long simulation time~$t$.}
   \label{fig:rw_steadystate}
\end{figure}

We use histograms to illustrate the random walk solution, density-normalised such that~$\rho_\text{bin} = c_\text{bin}/N_p/w_\text{bin}$ where~$c_\text{bin}$ and~$w_\text{bin}$ are the bin count and width respectively. We investigate the time step dependence of the membrane transit model for the simple domain described in \cref{sec:simple_domain}. \Cref{fig:rw_steadystate} shows random walk solutions using the new hybrid model and the reference transit model \cite{Fieremans2010}. We examine different time steps~$\del{t}$ (\SIlist{20;10;5;2;0.5;0.05}{\unitT}) and two solution times~\SIlist{20;1000}{\unitT}. For the short time solution, the reference transit model fails to preserve the initial steady-state solution near the barrier. As the time increases, a concentration imbalance develops across the barrier.
The difference in compartment density~$\dl.Delta.{U}$ across the membrane increases with~$\del{t}$ and appears to stabilise at some value as~$t \to \infty$: For~$\del{t} = \SI{0.05}{\unitT}$, the reference model shows an excess of concentration in the left compartment of~\SI{0.88}{\percent}, for~$\del{t} = \SI{20}{\unitT}$ this increases to~\SI{10.2}{\percent}. In \cref{sec:FluxAnalysis}, we mathematically prove the reference model limitations. On the other hand, the new hybrid model represents the expected solution with a constant~$U$ with a maximum of~$\dl.Delta.{U} = \SI{0.18}{\percent}$ between all values of~$t$ and~$\del{t}$, suggesting that there is no accumulation of walkers for any time step. The variation in bin densities for the interface model can be attributed to randomness of the simulation. At~$t = \SI{1000}{\unitT}$ the difference between observed $\rho_\text{bin}$~values relative to the expected~$\rho^*_\text{bin} = 1/\sum{L} = \num{0.025}$ has a median (among~$\del{t}$) standard deviation of~\num{0.0013}~(or \SI{5}{\percent} relative to~$\rho^*_\text{bin}$).

\subsection{Analysis of the transient-state}
\label{sec:rw_transient}

We perform two different transient-state analyses to independently investigate the influence of the step size and the influence of several permeability and diffusivity values.

\subsubsection{The influence of step size}
\label{sec:rw_error_stepsize}

\begin{figure}
    \centering
    \includegraphics{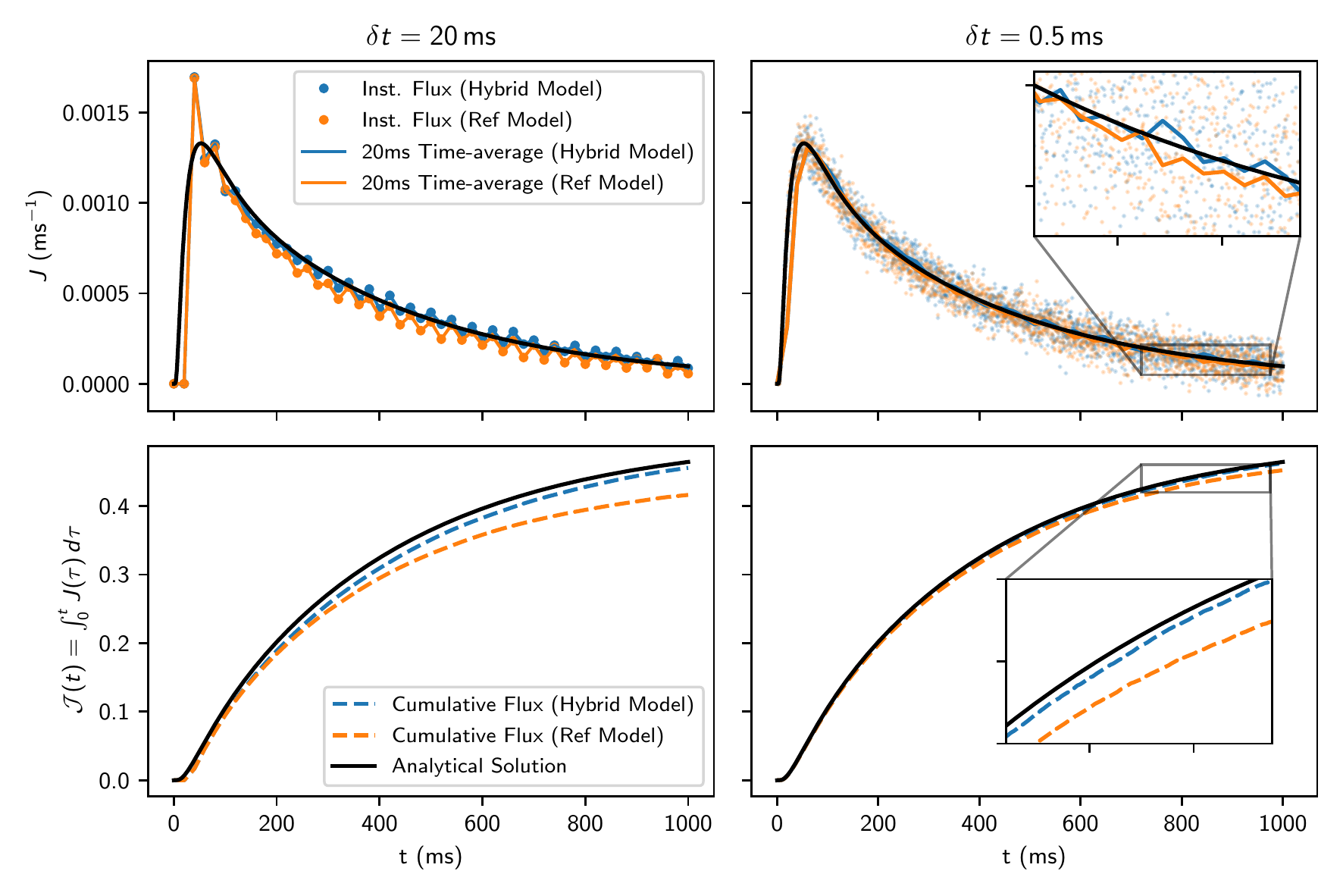}
    \caption{Top figures: Instantaneous and time-averaged (plotted as the average of every interval with $\dl.Delta.{t} = \SI{20}{\unitT}$) fluxes~$J(t)$ through the membrane as a function of simulation time. Bottom figure: Analytical and numerical net cumulative flux. We show results for two different step sizes~$\del{t}$:~\SIlist{20;0.5}{\unitT}. Domain and simulation parameters: $D_L = \SI{0.5}{\unitD}$, $D_R = \SI{2.5}{\unitD}$, $\kappa = \SI{0.05}{\unitK}$, $L = \SI{20}{\unitL}$, $N_p = \num{1e6}$.}
    \label{fig:rw_fluxes}
\end{figure}

Simulations are performed up to~$t = \SI{1000}{\unitT}$ using the largest time step ($\delta{t} = \SI{20}{\unitT}$) utilised in \cref{sec:rw_steadystate} and small time step of ~$\del{t} = \SI{0.5}{\unitT}$. \Cref{fig:rw_fluxes} shows the instantaneous and cumulative fluxes obtained from the random walk simulation at every time step alongside the analytical solution. For the random walk solution, we also plot a time-averaged flux over fixed intervals of~$\dl.Delta.{t} = \SI{20}{\unitT}$ to allow for comparison between both plots. Note that the time-average flux and instantaneous flux coincide for the largest time step. 
The (analytical) flux~$J$ rapidly increases early in the simulation and peaks at~$t = \SI{58.53}{\milli\second}$. As~$t$ increases, walkers continue to cross the membrane towards the initially empty compartment ($J > 0$ always). We observe that for a large time step ~$\del{t} = \SI{20}{\unitT}$ , the time-averaged/instantaneous flux for both transit models over-estimate the initial peak in magnitude. However, if we compare the cumulative fluxes at the final end-point of the simulation ($t=\SI{1000}{\unitT}$, the reference model underestimates the tail with a relative error of $-10.27\%$ compared to an error of $-1.87\%$ for the hybrid model. We observe that reducing the time step increases the number of walkers crossing the membrane increasing the overall convergence. For $\delta{t} = \SI{0.5}{\unitT}$, the relative error at the end of the simulation is reduced to $-2.6\%$ and to $-0.46\%$ for the reference and hybrid model respectively. This is consistent with the results observed in \cref{fig:rw_steadystate} for the steady-state analysis. 

\subsubsection{The influence of permeability and diffusivity}
\label{sec:rw_error_domainparam}

In order to study the effects of permeability and diffusivity, we consider the simple domain varying the ratio ($b=D_R/D_L$) between diffusivities in either side. We consider a constant left diffusivity of $D_L=\SI{2}{\unitD}$ and a range of ratios ($0.05<b<2.5$). We consider two permeability values (\SIlist{0.05;0.5}{\unitK}) that correspond to exchange times of the order of $50$Hz and $500$Hz respectively. The first permeability ($\SI{0.05}{\unitK}$) is linked to exchange times that are closer to what we would observe in human cells \cite{Seland2007} and the second permeability value covers higher exchange rates that might be useful for other applications. We perform simulations up to~$t = \SI{1000}{\milli\second}$ for six~different time steps~$\del{t}$: \SIlist{8;4;2;0.5;0.1;0.05}{\unitT} and 9 different varying diffusivity ratios \SIlist{2.5;1.8;1.6;1;0.4;0.2;0.1;0.05}. 
\begin{figure}
    \centering
    \includegraphics[width=0.86\textwidth]{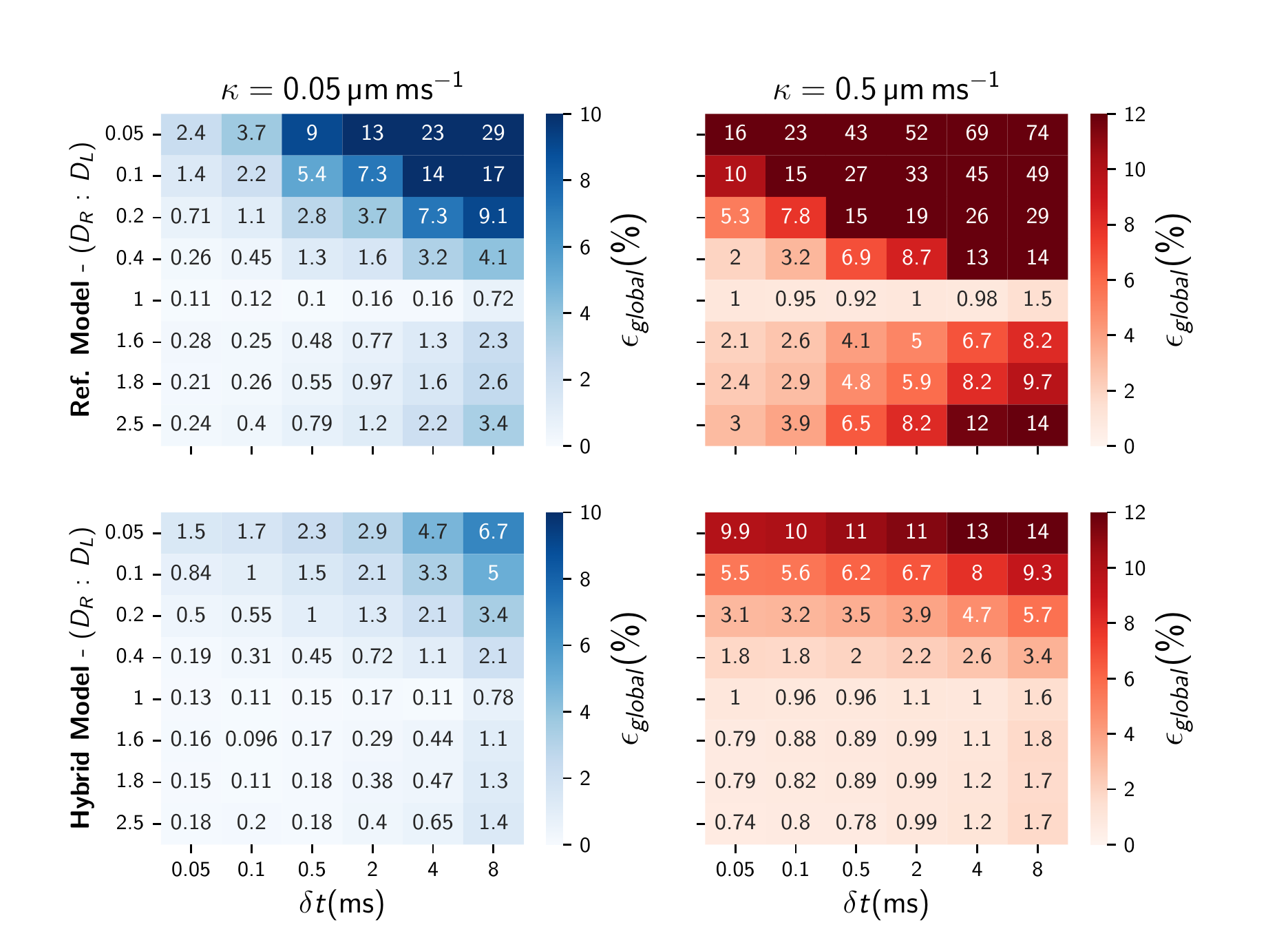}
    \caption{Errors in the random walk for different time steps~$\del{t}$ and permeabilities~$\kappa$ in four different domains (the left-most domain is the model domain used in \cref{sec:rw_steadystate,sec:rw_transient}). Simulations were run until~$t = \SI{1000}{\unitT}$ with $N_p = \num{1e6}$~walkers seeded at~$x_0 = \SI{10}{\unitL}$. Top: The global error (relative to~$\int{\mathcal{J}_\text{ana}\dl2{t}}$) quantifies the degree to which the solution fluctuates around the analytical solution during the simulation. Bottom: The cumulative error (relative to~$\mathcal{J}_\text{ana}$) represents the error in net flux at the end of the simulation.}
    \label{fig:rw_errorstudy}
\end{figure}
\Cref{fig:rw_errorstudy} shows the global relative errors that have been computed by evaluating the integral difference between the numerical and analytical cumulative flux. The top plots and the bottom plots illustrate the global errors for the reference model and hybrid model respectively. As mentioned in \cref{methods_transient_steady_simple}, each simulation has been performed initialising the walkers in a partial uniform region within the left compartment. For a constant permeability, we observe that the global errors escalate as we increase the difference between diffusivities. These increments are consistently lower for the hybrid model as it includes the probability when transitioning between two different media through \cref{eqn:pt_Maruyama}. Similarly to \cref{fig:rw_fluxes}, lowering the step size increases the number of walker collisions with the membrane resulting in an overall faster convergence and accuracy. As it can be observed in \cref{eqn:pt_hybridmodel}, the hybrid model incorporates the reference model to solve the membrane/interface probability. In \cref{sec:FluxAnalysis_interfacecondition}, we show that the reference model does not preserve the interface reflection and leads to a permeability-related error. As a result, both models show a permeability error dependency, however, the hybrid model relative errors are consistently lower due to the initial error reduction in the diffusion media.
\subsection{Transient and steady-state solutions in the histology-based domain}
\label{sec:transient_histology}
\begin{figure}
    \centering
    \includegraphics{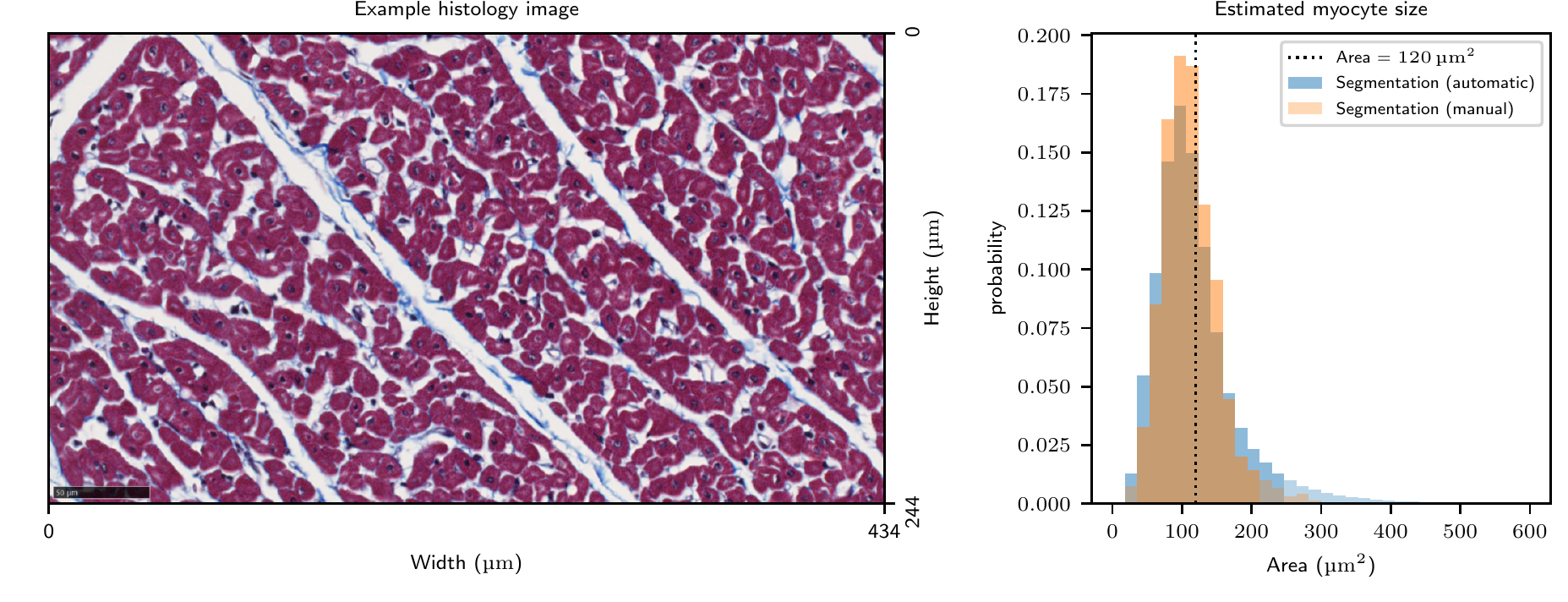}
    \caption{Illustration of the process of synthesising a 1D~domain from histology data. Left Figure: An example of a region of histology from a wide-field microscopy image. This is part of a large vertical stack of histological slices obtained from the mesocardium of swine. Cardiomyocytes (red-purple) are cut perpendicular to the imaging plane. Extra-cellular space is white, while collagen is stained blue. Right Figure: Distribution of cell sizes from automatic segmentation for the entire stack of images as well as manual labelling of a small representative region.}
    \label{fig:histology_domain}
\end{figure}
\begin{figure}
    \centering
    \includegraphics{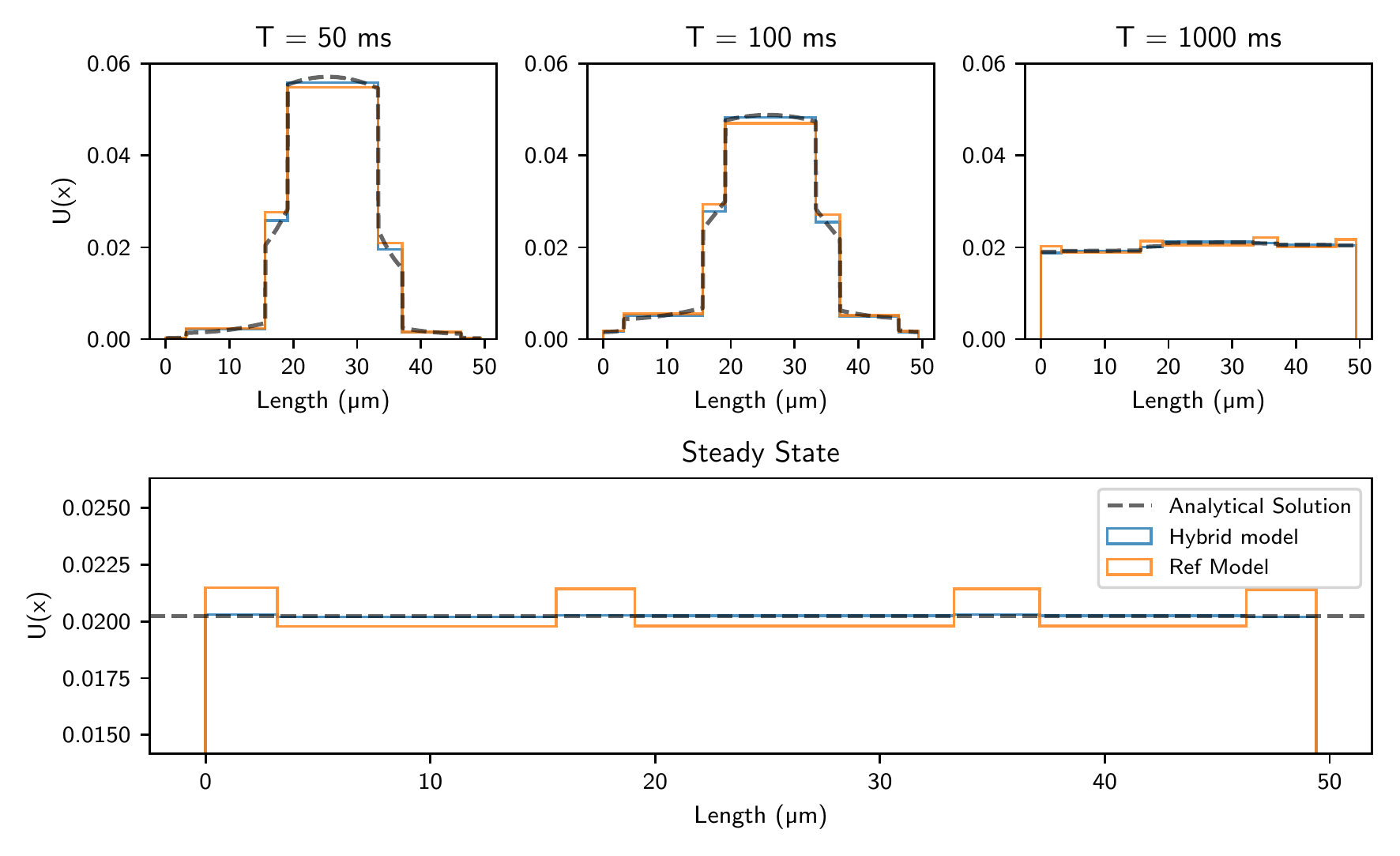}
    \caption{Analytical and random walk solutions (using $N_p=\num{1e6}$ walkers and a time step of $\del{t}=\SI{1.5}{\unitT}$) at $t=\SI{1000}{\unitT}$. Top: Transient solution~$U(x, t; x_0)$ for initial concentration at $x_0$~located in the centre of the domain. Bottom: Steady-state solution after uniform seeding of walkers in the domain. The reference model results in accumulation of walkers in ECS compartments. Note that this effect is visually amplified by the choice of axis data range.}
    \label{fig:states_complexdomain}
\end{figure}

\Cref{fig:histology_domain} illustrates the distribution of cell sizes for the manual and automatic segmentation~\cite{ISMRM2018:Rose}. As explained in \cref{sec:methods_synthesised_domain}, we recreate a histology-based domain by considering circular cross-sections for cardiomyocytes. The segmentation data shows a mean cross-sectional area of~$\mu = \SI{120}{\unitL\squared}$ and a standard deviation of~$\sigma = \SI{40}{\unitL\squared}$. We utilise these parameters to create a synthesised histology-based domain using a normal distribution (~$\mu \pm 2 \sigma$) for the intra-cellular space and a uniform distribution ($\SI{3}{\unitL}$-$\SI{5}{\unitL})$ for the extra-cellular compartments. The resulting domain is recreated by drawing both intra-cellular and extra-cellular distributions until reaching a total length of $\SI{49.5}{\micro\meter}$. \Cref{fig:states_complexdomain} shows the final steady state and three different transient states for the histology-based domain. The transient states are analysed at three different diffusion times $T=50,100,1000 ms$ considering that all the particles are initialised in the middle of the domain $X_{0}=\SI{24.75}{\micro\meter}$. We observe good agreement between the hybrid model and the analytical solutions. We notice that the reference model tends to be less restrictive in the initial transient states leading to higher concentrations in the ECS. This accumulation of walkers is present and is carried during the entire simulation until reaching the steady-state. This excess of walkers in the ECS persists throughout the simulation and is consistent with the findings in \cref{sec:rw_steadystate} and \cref{sec:rw_transient}.
\subsection{DWI results}
\begin{figure}
    \centering
    \includegraphics{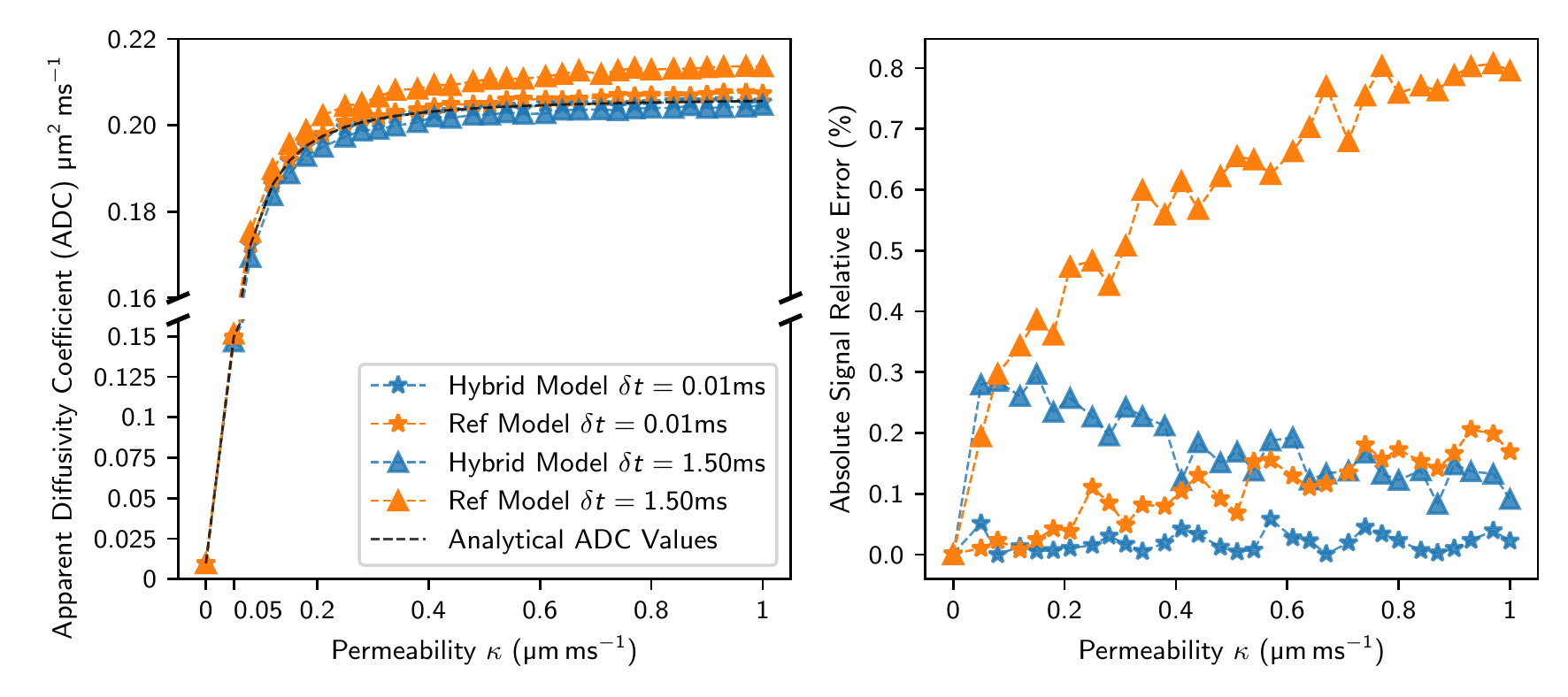}
    \caption{Absolute relative signal error and ADC values for a wide range of permeability values considering a very small ($\delta{t}=\SI{0.01}{\unitT}$) and large time step ($\delta{t}=\SI{1.5}{\unitT}$). All the simulations consider $N_p=10^6$ walkers, $D_{ECS}=\SI{2}{\unitD}$ and $D_{ICS}=\SI{0.5}{\unitD}$}
    \label{fig:error_signal}
\end{figure}

We analyse how the diffusion errors observed in \cref{sec:transient_histology} affect the DWI signal and apparent diffusion coefficient (ADC). We compare the signal attenuation and ADC with the theoretical analytical results. As mentioned in \cref{methods:application_dwi}, we seed the walkers uniformly and consider the narrow pulse approximation (NPA). \Cref{fig:error_signal} shows the ADC and the absolute signal relative error for a large interval of membrane permeabilities and two different step sizes. \Cref{eqn:NPA_mcrw} shows the relation between the numerical signal error and the displacement of the walkers. Similarly to the flux analysis, we notice a strong dependency between the permeability and the accuracy of both transit models. If we compare the errors for different increasing permeability values, we observe that the reference model progressively becomes less accurate. We initially report low relative errors ($\approx 0.3\%$) for $\delta{t}=1.5\si{\unitT}$ that scale up to $0.8\%$ for the reference model. These signal and ADC errors can be reduced when decreasing $\delta{t}$ due to the increment of particle collisions against the membranes. In the application that we are interested in, the cardiomyocytes have low permeability values ($\kappa <0.05 \si{\unitK})$ \cite{Sobol1990} where both transit models perform similarly with low relative errors.
\clearpage
\section{Discussion}

Simulating the transit of particles through a semi-permeable membrane separating media with differing diffusivities can be achieved using analytical or numerical techniques. In this work, we present novel methods for determining the particle density in a 1D~domain and the flux through a membrane.

Monte Carlo random walk based method are frequently used to solve the diffusion equation~\labelcref{eqn:diffusion} in a range of applications. When a walker encounters a barrier such as a semi-permeable membrane, the probability of transit needs to be accurately computed and the walker step size must be adjusted to satisfy \cref{eqn:afterStep} when transiting between compartments of differing diffusivities~\cite{Szafer1995}. One of these so-called transit models is the popular method described in \cref{eqn:pt_Fieremans}~\cite{Fieremans2010}. As described in the original paper, this reference transit model~\cite{Fieremans2010} requires a highly resolved time step. For this reason, it is suggested that the maximum time step ($\delta{t}_{max}$) is limited by a maximum possible transmission probability of $0.01$~\cite{Fieremans2010}. This can be a computational challenge for long and even short diffusion time simulations when other Monte Carlo parameters such as the number of walkers~$N_p$ or the number of unique experiments need to be considered. In this work we propose a new hybrid transit model with the motivation of lifting this restriction. The new model that we present is based on treating the membrane as an infinitesimal space such that the diffusivity gradient and the membrane permeability can be considered as two independent factors. It is important to note that the hybrid model is built on top of the reference model as it is used to solve the membrane probability. Using an analytical solution as a gold standard, we have studied the accuracy of the hybrid and reference transit model~\cite{Fieremans2010} when exceeding the maximum step size ($\delta{t}_{max}$). We have assessed both models by analysing the membrane flux varying several parameters in a simple geometry domain.  

We have found that the reference model leads to errors in the net migration of walkers and results in concentration imbalances for steady-state solutions. Further analysis in \cref{sec:FluxAnalysis} demonstrated that the transit model inherently tends to accumulate walkers when the membrane divides two compartments of different diffusivities. In \cref{sec:FluxAnalysis_interfacecondition} we also note that the interface reflection condition in \cref{eqn:interface_reflection} is not respected as~$\kappa \to \infty$ and this is consistent with the findings observed for both transit models in \cref{fig:rw_errorstudy} and with the relation between diffusivity/permeability to fit within the maximum time step ($\delta{t}_{max}$).

The numerical simulations performed in this study consider a fixed step size. In \Cref{sec:rw_transient}, we have considered a partial uniform distribution of walkers in the left compartment to avoid sampling error for large step sizes due to the unique number of possible walker positions. From the numerical simulations in the simple domain, we numerically prove the limitations of the reference model for varying diffusivities. From our steady-state and transient findings, we conclude that these imbalances increase for higher step sizes and longer diffusion times. Further numerical analysis shows that the hybrid model substantially reduces the diffusivity and permeability-related errors present in the reference model. It is important to note that when there is no diffusivity discontinuity the hybrid model is in pure essence the reference model itself. Our results demonstrate that the hybrid model successfully lifts the time step restriction and other numerical limitations. 

One important application of the methods presented in this work is in the simulation of diffusion weighted imaging~(DWI). Previous numerical simulations of diffusion tensor imaging~(a variant of DWI) in the heart were based on impermeable membranes~\cite{Rose2019} and resulted in more anisotropic diffusion than commonly found in imaging studies~\cite{vonDeuster2016}. Initial work using finite volume methods~\cite{ISMRM2019:Rose} has suggested that this underestimation of anisotropy is likely due, at least in part, to the failure to include permeability within the model.

The simulations in \cref{sec:transient_histology} use a domain constructed based on microscopy images of histology sections. Using a previously developed method~\cite{ISMRM2018:Rose} to automatically segment cells, we have extracted characteristic cell sizes to generate a representative one-dimensional domain. A mean cell area of~\SI{120}{\unitL}, which corresponds to a diameter of~\SI{12.4}{\micro\meter} assuming circular cells was obtained from segmentation of the data. This value is slightly lower than the reported range of~\SIrange{17}{25}{\micro\meter} in~\cite{Tracy2011}. However, tissue is known to shrink during histological preparation and this may be compensated for during image processing by morphing the domain as shown in~\cite{Rose2019}.

The cardiomyocyte membrane permeability ($\kappa=\SI{0.05}{\unitK}$) and the diffusivity values ($D_\text{ECS}=\SI{2}{\unitD}$, $D_\text{ICS}=\SI{0.5}{\unitD}$) in the histology-based simulations limit the $dt_{max}$ to extremely low values $\delta{t} \approx \SI{0.002}{\milli\second}$. Similarly to the simple domain analysis, our histology-based results for the reference model show accumulation of particles in the ECV that are carried along the DWI signal computation. We have observed that both models have very small and similar errors for low permeability in the range of cardiomyocytes. We have shown increasing errors in the results of the reference model with larger permeability values. The findings regarding the limitations of the reference model at high~$\kappa$ and larger ratios of~$D$ between compartments are particularly important for applications beyond biological tissue such as heat transfer where different sets of parameters may be required. For example, a thin layer of material in heat conduction problems can be modelled as a membrane with permeability~$D/L$.

The complex three-dimensional cardiac microstructure is often successfully modelled as being two-dimensional. We have attempted to reduce it further to a 1D~domain to make use of analytical techniques to validate our new hybrid model and compare the performance with a previously random walk transit model.. This has provided useful insights into the behaviour of both model for multiple applications. In future work, we plan to extend the thorough validation performed here into higher dimensions. For example, there exists an analytical solution in~2D using radial coordinates~\cite{Singh2008} and the finite volume method offers a well-understood reference solution~\cite{ISMRM2019:Rose}. Computationally, the extension of the random walk into higher dimensions is trivial by considering the boundary-normal distance for the distance~$\del{x}_i$ in \cref{eqn:pt_Fieremans} as discussed in~\cite{Fieremans2010}. The interface model in \cref{eqn:pt_Maruyama} is valid for any dimension as it relies on the \emph{length} of the step~\cite{Maruyama2017}. Yet other methods involving kinetic approximation~\cite{Lejay2010} or a dual-probability-Brownian motion scheme~\cite{Gong2020} warrant further investigation.

\section{Conclusions}

Modelling diffusion within non-simple layered media requires a correct treatment of the particle transit at membranes when numerical solutions are employed.

We investigate the accuracy of a previously proposed transit model for varying diffusion coefficients in the presence of permeable barriers. We compared the numerical results to an analytical method and show the limitation of this reference transit model. We propose an alternative to this problem by considering a hybrid transit model that treats the transitioning between different media and the membrane as two separate probabilities. By comparing numerical and analytical simulations, we conclude that the new transit model performs with lower errors and successfully lifts the step restriction reducing the accumulation of walkers observed in the reference model. While random walk methods for the solution of the diffusion equation in layered media have a number of potential applications, we have considered simulating diffusion-weighted imaging in cardiac tissue.

For the given range of extracellular compartment lengths we find that the time step is already very restricted by the domain itself as the walkers cannot cross multiple compartments. We conclude that in application to cardiac tissue, both transit models present very small and similar errors in the apparent diffusion coefficient, an important integral quantity for diffusion-weighted imaging. However, other applications where the compartment length  are larger and both permeabilities and diffusivites are higher, the potential of this new hybrid model can be fully realised. In these cases, the hybrid model becomes a computationally more efficient and more accurate alternative for solving the diffusion equation than existing models.

\section*{Acknowledgments}

Confocal microscopy images were obtained in collaboration with Dr.~Padmini Sarathchandra at The Magdi Yacoub Institute, and the Facility for Imaging by Light Microscopy at Imperial College London.
Dr.~Sonia Nielles-Vallespin kindly provided the wide-field microscopy images of histology slices.

\section*{Declarations}

\section*{Author contributions statement}
\textbf{Conceptualization}: IA, JR, DJ.D; \textbf{Formal analysis and investigation}: IA, JR; \textbf{Writing - original draft preparation}: IA, JR ; \textbf{Writing - review and editing}: AD.S, DJ.D; \textbf{Funding acquisition}: AD.S, DJ.D ; \textbf{Supervision}: AD.S, DJ.D; \textbf{Software}:IA,JR

\section*{Additional information}
\textbf{Funding}
This work was funded by British Heart Foundation grants RE/13/4/30184 and RG/19/1/34160.
\noindent
\textbf{Conflict of interest}
The authors declare that they have no conflict of interest.
\noindent
\bibliography{bibliography}

\begin{appendices}
\section{Reference transit model derivation}
\label{sec:ModelDerivation}

\begin{figure}
    \centering
    \includegraphics{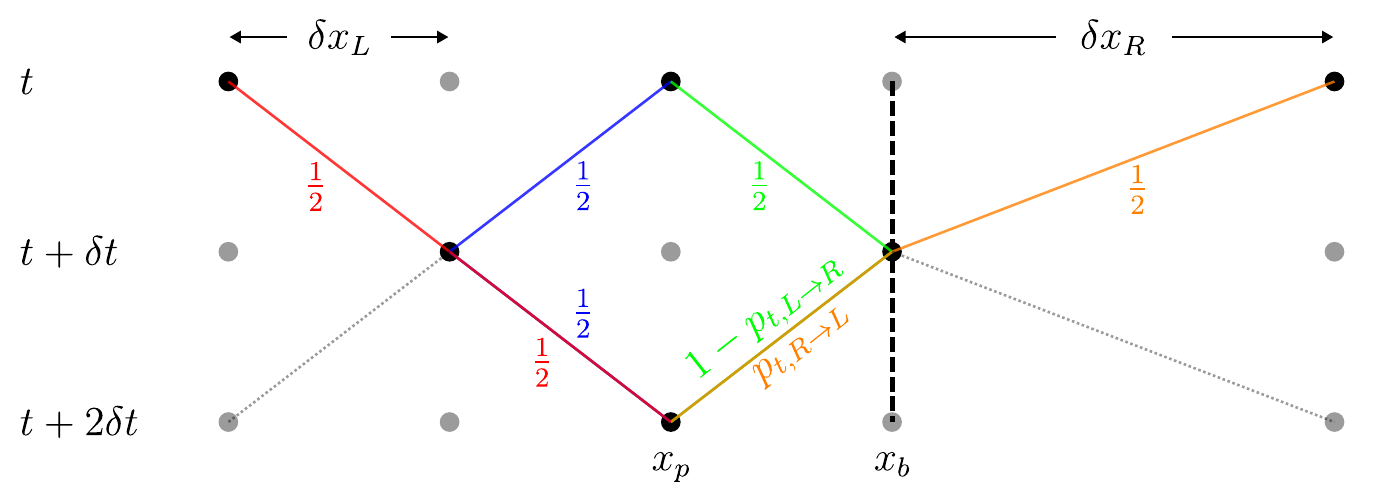}
    \caption{A grid of walker positions~$x$ near a permeable barrier located at~$x_b$. The concentration at~$x_p$, i.e.~$U(x_p, t + 2\del{t})$, is composed of the contribution of three different walker positions at time~$t$ (two time steps prior) through four different paths. The green path reflects at the barrier with probability~$1 - p_{t, L \to R}$, while the orange path passes through the barrier with transit probability~$p_{t, R \to L}$.}
    \label{fig:Powles_probabilities}
\end{figure}

The work in~\cite{Powles1992} presents a derivation of the transit probability~$p_t$ for walker positions on a lattice based on a constant step length in the entire domain. Here, we consider domains where the diffusivity between compartments may differ and re-derive the probability of transit for the domain in \cref{fig:Powles_probabilities} considering the different step lengths.

Consider the probability of finding a walker at~$x_p$ as the sum of probabilities of neighbouring walkers jumping to~$x_p$. Away from interfaces, this results in
\begin{equation}
    U(x_p, t+\del{t}) = \frac{1}{2} U(x_p-\del{x}, t) + \frac{1}{2} U(x_p+\del{x}, t) \, .
\end{equation}
Selecting~$x_p$ to be near a barrier at~$x_b$ (see \cref{fig:Powles_probabilities} for the possible paths of walkers to reach this position) results in additional terms involving~$p_t$ (both~$p_{t, L \to R}$ and~$p_{t, R \to L}$). The barrier location is chosen to coincide with a grid point. A walker located here will proceed left/right according to the appropriate probability of transit. The contributions of different $U(x, t)$~terms to~$U(x_p, t + 2\del{t})$ in \cref{fig:Powles_probabilities} (with the path colour indicated) are:
\begin{equation}
    U(x_p, t + 2\del{t})
    =
    \underbrace{
    \frac{1}{4} U(x_p - 2\del{x}_L, t)
    }_\text{free diffusion (red)}
    +
    \underbrace{
    \frac{1}{4} U(x_p, t)
    }_\text{free diffusion (blue)}
    +
    \underbrace{
    \frac{1 - p_{t, L \to R}}{2} U(x_p, t)
    }_\text{reflection (green)}
    +
    \underbrace{
    \frac{p_{t, R \to L}}{2} U(x_p + \del{x}_L + \del{x}_R, t)
    }_\text{transit (orange)}
\end{equation}

One may now express~$x_p$ relative to~$x_b$, e.g. the contribution of the orange path from the right side of the barrier is:
\begin{equation}
    U(x_p + \del{x}_L + \del{x}_R, t) = U(x_b + \del{x}_R, t)
\end{equation}
%
This allows us to perform a Taylor series expansion of every term around~$x_b$, taking care to expand infinitesimally to the left and right of the barrier as appropriate.
Following that, we apply the diffusion equation~\labelcref{eqn:diffusion} to remove the time derivative. The boundary conditions
in \cref{eqn:leather_BC,eqn:interface_reflection,eqn:afterStep} as well as \cref{eqn:flux_BC} and its extension
\begin{equation}
    D_R \diffp[2]{U}{x}[R] = D_L \diffp[2]{U}{x}[L]
\end{equation}
allow us to express everything in terms of $\left.U\right|_L$, $\diffp{U}/{x}[L]$, and $\diffp[2]{U}/{x}[L]$ as well as $p_{t, L \to R}$ and $\del{x}_L$. Since we did not specify any constraints regarding the values of~$D_L$ and~$D_R$ or their relationship, $p_t$ is equivalent to~$p_{t, i \to j}$ in \cref{eqn:pt_Fieremans} given the diffusivities~$D_L=D_i$ and~$D_R=D_j$.

The final expression is:
\begin{equation}
    0
    =
    2 p_t \left( 1 - \sqrt{\frac{D_L}{D_R}} \right) \left.U\right|_L
    +
    2 \left( \left( 1 - \left( 1 + \frac{D_L}{D_R} \right) p_t \right) \del{x}_L - p_t \sqrt{\frac{D_L}{D_R}} \frac{D_L}{\kappa} \right) \diffp{U}{x}[L]
    + p_t \left( 1 - \sqrt{\frac{D_L}{D_R}} \right) \del{x}_L^2 \diffp[2]{U}{x}[L]
\end{equation}
It is important to recognise that this recovers the expression derived by Powles et al. in the case $D_L = D_R$~\cite{Powles1992}. It also suggests that the model in \cref{eqn:pt_Fieremans} as derived in~\cite{Fieremans2010} relies on ignoring the terms associated with~$U$ and~$\diffp[2]{U}/{x}$. Since we cannot make statements about the magnitude of these terms, the former omission is only valid provided~$p_t \to 0$ or~$D_L/D_R \to 1$, while the latter requires $\del{x} \to 0$ as well. These relations are in line with observations of the model's behaviour throughout this work.

\clearpage

\section{Hybrid model derivation}
\label{S_Appendix_hybridmodel}
First we consider the case of the membrane placed on the ``slow'' side, where the diffusion coefficient $D_2$ is lower.
If the walker attempts a crossing from the slow side, it first transits through the membrane with probability $P_{B_2}$ or is reflected with $1 - P_{B_2}$.
It always transits through the step change in diffusion coefficient, because it enters a region of higher diffusion coefficient.
If the walker originates on the fast side, it first needs to pass the step change in diffusion coefficient with probability $P_D$.
Subsequently the membrane on the slow side either permits transit with $P_{B_2}$, which terminates the interaction, or reflects it with $1 - P_{B_2}$, which causes the walker to attempt and always succeed in transiting through the step change in diffusion coefficient.
This can be summarised with the two probabilities
\begin{subequations}\label{eqn:probabilities_slow}
\begin{align}
    P_{\textnormal{fast} \to \textnormal{slow}} &= P_D P_{B_2}
    \, ,
    \\
    P_{\textnormal{slow} \to \textnormal{fast}} &= P_{B_2}
    \, .
\end{align}
\end{subequations}

\begin{figure}
    \centering
    \includegraphics[angle=90,origin=c,width=0.6\textwidth]{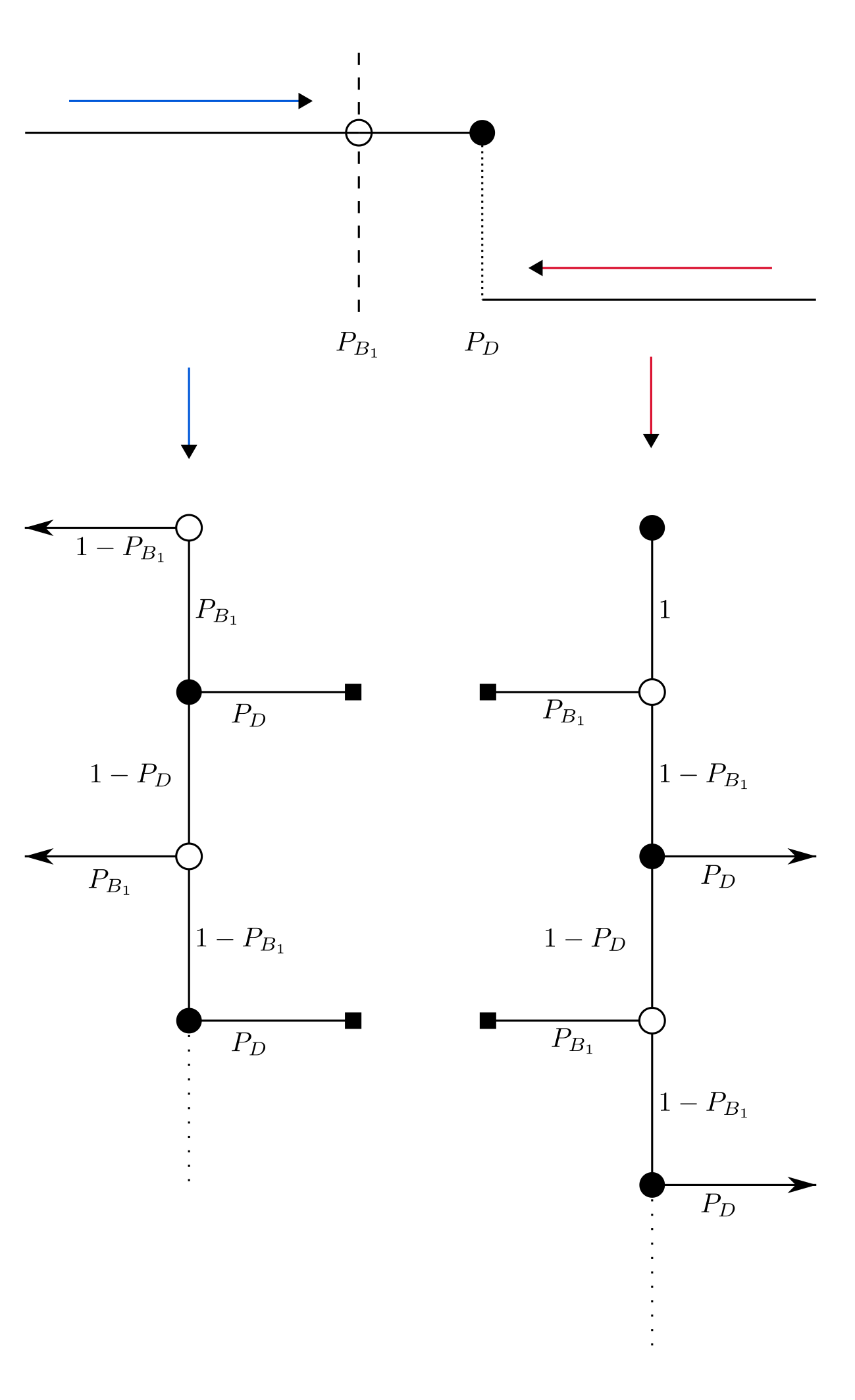}
    \vspace{-5em}
    \caption{Interface model for a membrane on the fast side}
    \label{fig:interface_fast}
\end{figure}

An alternative to the above model is to place the membrane on the ``fast'' side, where the diffusion coefficient $D_1$ is higher.
This scenario results in walkers reflecting between the two interfaces continuously until eventually exiting on either side.
The diagram in \cref{fig:interface_fast} illustrates this.
By summing the probabilities of all different exit conditions, we obtain
\begin{subequations}\label{eqn:probabilities_fast}
\begin{align}
    p_{\textnormal{fast} \to \textnormal{slow}} &= \sum_{i=1}^{\infty}{ P_{B_1} \left( 1 - P_D \right)^{i-1} \left( 1 - P_{B_1} \right)^{i-1} P_D }
    \label{eqn:p_fast_fast2slow}
    \, ,
    \\
    p_{\textnormal{slow} \to \textnormal{fast}} &= \sum_{i=1}^{\infty}{ \left(1 - P_{B_1} \right)^{i-1} \left( 1 - P_D \right)^{i-1} P_{B_1} }
    \label{eqn:p_fast_slow2fast}
    \, .
\end{align}
\end{subequations}
Here, the infinite sums only converge if $|(1-P_{B_1}) (1-P_D)| < 1$, which is the case given that all $P \leq 1$.
The probabilities without power of $i-1$ can be removed from the summation and, knowing that
\begin{equation}
    \lim_{n \to \infty}{ \sum_{i = 0}^{n}{ (1-A)^i (1-B)^i } } = \frac{1}{A + B - AB} \, ,
\end{equation}
we obtain the following expressions for ...
\begin{subequations}
\begin{align}
    p_{\textnormal{fast} \to \textnormal{slow}} &= \frac{P_{B_1} P_D}{P_{B_1} + P_D - P_{B_1} P_D}
    \, ,
    \\
    p_{\textnormal{slow} \to \textnormal{fast}} &= \frac{P_{B_1}}{P_{B_1} + P_D - P_{B_1} P_D}
    \, .
\end{align}
\end{subequations}
To proof that both methods are equivalent we need to show that the following equality is valid
\begin{equation}
    \frac{P_{B_2}}{P_{B_1}} (P_{B_1} + P_D + P_{B_1}P_D) = 1
    \label{eq:ratio_eq}
\end{equation}
From \cite{Fieremans2010} we know that 
\begin{equation}
    P_{B_i} = \dfrac{\dfrac{2\sqrt{2\delta{t}}\kappa}{\sqrt{D_i}}}{1+\dfrac{2\sqrt{2\delta{t}}\kappa}{\sqrt{D_i}}} = \dfrac{\dfrac{a}{\sqrt{D_i}}}{1+\dfrac{a}{\sqrt{D_i}}}
    \label{eq:pb_developed}
\end{equation}
Note that $a$ is a constant. If we substitute $P_{B_1}$ and $P_{B_2}$ in the right hand side of \cref{eq:ratio_eq} for \cref{eq:pb_developed} and $P_D$ for \cref{eqn:pt_Maruyama} we arrive at the following simplified equation:
\begin{equation}
    \dfrac{1 + \dfrac{a}{\sqrt{D_1}}}{1 + \dfrac{1}{\sqrt{D_2}}}
    +
    \dfrac{\dfrac{a}{\sqrt{D_1}}}{1 + \dfrac{1}{\sqrt{D_2}}} \dfrac{\sqrt{D_1}}{\sqrt{D_2}}
    -
    \dfrac{\dfrac{a}{\sqrt{D_1}}}{1 + \dfrac{1}{\sqrt{D_2}}}
    \label{eq:substitute_probabilities_hybridmodel}
\end{equation}
Doing some basic algebraic operations it is easy to prove that \cref{eq:substitute_probabilities_hybridmodel} is reduced to $1$ validating the equality in \cref{eq:ratio_eq} and showing that both methods give the same probability.

\section{Flux analysis}
\label{sec:FluxAnalysis}

\begin{figure}
    \centering
    \includegraphics{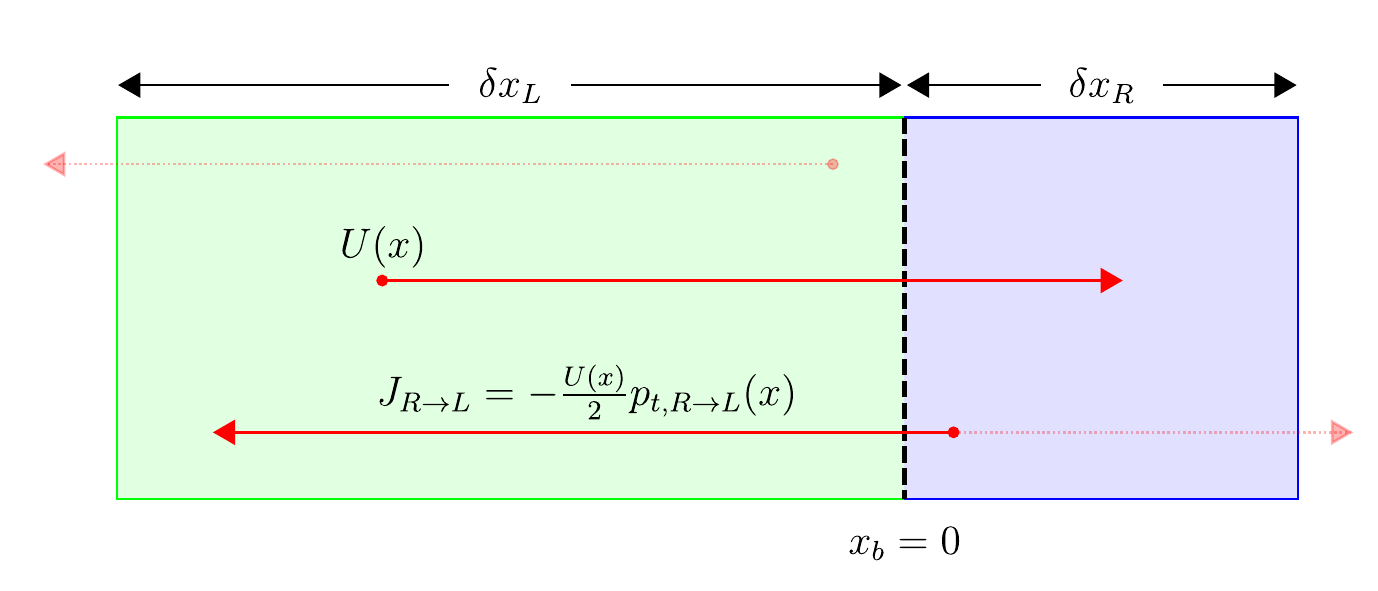}
    \caption{A graphical explanation of the flux analysis around two compartments~($D_L > D_R$) separated by a permeable barrier with asymmetric transit probabilities~$p_{t, L \to R}$ and~$p_{t, R \to L}$. The control volume is bounded by the step lengths~$\del{x}$ on either side. Walkers located inside this volume at a point with concentration/density~$U(x)$ add to the flux component~$J$ with their weighted contribution~$\pm U(x) p_t(x)$, provided they step towards the barrier (accounted for by the factor~$\sfrac{1}{2}$).}
    \label{fig:fluxanalysis}
\end{figure}

Here, we further analyse the steady-state case and offer an explanation for the differences in the models' behaviour observed in~\cref{sec:rw_steadystate} and elsewhere in this work. Consider the control volume around a barrier bounded by the maximum step length~$\del{x}$ on either side as illustrated in~\cref{fig:fluxanalysis}. Note the difference between the running variable~$x$~(and its differential~$\dl{x}$) and the finite step length~$\del{x}$.

The net flux~$\vector{J} = -D\gradient{U}$ can be split into its individual components, i.e.~the amount of walkers/concentration crossing from left to right or right to left, whose magnitudes are given by:
\begin{subequations}
\begin{align}
    J_{L \to R} &= \frac{1}{\del{t}} \int_{-\del{x}_L}^{0}{ +\frac{U(x)}{2} p_{t, L \to R}(x) \dl2{x} } \, , \\
    J_{R \to L} &= \frac{1}{\del{t}} \int_{0}^{ \del{x}_R}{ -\frac{U(x)}{2} p_{t, R \to L}(x) \dl2{x} } \, .
\end{align}
\end{subequations}
The integral bounds on the two sides cover the farthest that a walker can be located away from the boundary on either side in order for its step~$\del{x}$ to interact with the barrier. While the concentration density~$U(x)$ in general depends on position~$x$, we ignore this dependency here because we assume~$U = \text{constant}$ in the domain for the steady-state case. The factor~$\sfrac{1}{2}$ accounts for the fact that only half the walkers are expected to step towards the barrier. The $\pm$~signs are included to designate the directionality of the flux components~$J$ such that~$\sum{J} = 0$. For simplicity, we now omit this and use \emph{magnitudes} only. We also cancel the terms~$U$ and~$\del{t}$ and henceforth use the probability integrals~$F$:
\begin{equation}
    F_{A \to B} = \int_{}^{} p_{t,A \to B}(x) \dl2{x} \, .
\end{equation}

The steady-state solution requires the net flux~$\sum{J} = 0$, i.e.~$F_{L \to R} = F_{R \to L}$. If not, an imbalance in~$U(x)$ will develop. If the diffusivities on both sides are equal~($D_L = D_R = D$), the models considered in this work reduce to~$p_{t, L \to R} = p_{t, R \to L}$. Furthermore, by definition~$\del{x}_L = \del{x}_R$ and hence~$\sum{J} = 0$. Next, we consider the different models for the non-trivial case~$D_L \neq D_R$.

\subsection{Interface model}

In the case of an infinitely permeable membrane, we consider the interface model in~\cref{eqn:pt_Maruyama}.
The probability of transit for interfaces with discontinuous diffusivity~$D$ (ignoring permeability~$\kappa$) is given by
\begin{subequations}
\label{eqn:pt_Maruyama_2sided}
\begin{align}
    p_{t, L \to R} &= \min\left( 1, \sqrt{\frac{D_R}{D_L}} \right) \, , \\
    p_{t, R \to L} &= \min\left( 1, \sqrt{\frac{D_L}{D_R}} \right) \, .
\end{align}
\end{subequations}

For this model, the probabilities~$p_t$ do not depend on~$x$ but only on~$D$. As a result, we can remove~$p_t$ from the integral and simplify the result:
\begin{subequations}
\begin{align}
    F_{L \to R} &= p_{t, L \to R} \int_{-\del{x}_L}^{0} \dl2{x} = p_{t, L \to R} \del{x}_L = p_{t, L \to R} \sqrt{2 D_L \del{t}} \, , \\
    F_{R \to L} &= p_{t, R \to L} \int_{0}^{ \del{x}_R} \dl2{x} = p_{t, R \to L} \del{x}_R = p_{t, R \to L} \sqrt{2 D_R \del{t}} \, .
\end{align}
\end{subequations}
To remove the awkward $\min$-operator for~$p_t$, consider the two distinct cases in \cref{tab:maruyama_cases} separately. For both combinations of $p_t$~values it is clear that~$F_{L \to R} = F_{R \to L}$. In fact,~$F = \sqrt{2 D_\text{min} \del{t}}$ which we also observe in numerical simulations with~$\kappa = \infty$.

\begin{table}[!h]
    \centering
    \caption{The two cases of~$D_L$ and~$D_R$ considered in \cref{eqn:pt_Maruyama_2sided} and their corresponding probabilities of transit~$p_t$ and resulting fluxes~$F$.}
    \label{tab:maruyama_cases}
    \begin{tabular}{c||c|c|c}
        $D$ & $p_{t, L \to R}$ & $p_{t, R \to L}$ & $F$ \\
        \hline
        $D_L > D_R$ & $\sqrt{\frac{D_R}{D_L}}$ & $1$ & $\sqrt{2 D_R \del{t}}$ \\
        $D_L < D_R$ & $1$ & $\sqrt{\frac{D_L}{D_R}}$ & $\sqrt{2 D_L \del{t}}$ \\
    \end{tabular}
\end{table}

\subsection{Membrane model}

The authors in~\cite{Fieremans2010} introduce two probabilities of transit~$p_t$, going from~$L$ to~$R$ and from~$R$ to~$L$:
\begin{subequations}
\label{eqn:pt_Fieremans_2sided}
\begin{align}
    p_{t, L \to R} &= \frac{2 \kappa s_L}{D_L + 2 \kappa s_L} \approx \frac{2 \kappa s_L}{D_L} \, , \\
    p_{t, R \to L} &= \frac{2 \kappa s_R}{D_R + 2 \kappa s_R} \approx \frac{2 \kappa s_R}{D_R} \, .
\end{align}
\end{subequations}
Each~$p_t$ depends on the diffusivity~$D$ of the original compartment and the distance~$s$ from the step origin to the barrier. The latter is equal to~$\mp x$ since~$x_b=0$, such that~$s_L \in [-\del{x}_L, 0]$ and~$s_R \in [0, \del{x}_R]$ respectively. Note that the simplification on the right-hand sides of \cref{eqn:pt_Fieremans_2sided} only applies if~$2 \kappa s \ll D$. We will analyse both cases below.

\subsubsection{Larger time step}

Using the full, non-approximated term for~$p_t$ from \cref{eqn:pt_Fieremans_2sided}, we obtain the following:
\begin{subequations}
\begin{align}
    F_{L \to R} &= \int_{-\del{x}_L}^{0} p_{t, L \to R}(x) \dl2{x} = \int_{-\del{x}_L}^{0} \frac{x}{x - \frac{D_L}{2 \kappa}} \dl2{x} \, , \\
    F_{R \to L} &= \int_{0}^{ \del{x}_R} p_{t, R \to L}(x) \dl2{x} = \int_{0}^{\del{x}_R} \frac{x}{x + \frac{D_R}{2 \kappa}} \dl2{x} \, .
\end{align}
\end{subequations}
The integrals can be solved to give:
%
\begin{subequations}
\begin{align}
    F_{L \to R}
    &
    = \del{x}_L + \frac{D_L}{2 \kappa} \left( \ln\left( -\frac{D_L}{2 \kappa} \right) - \ln\left( -\del{x}_L - \frac{D_L}{2 \kappa} \right) \right)
    \, ,
    \\
    F_{R \to L}
    &
    = \del{x}_R + \frac{D_R}{2 \kappa} \left( \ln\left( \phantom{-}\frac{D_R}{2 \kappa} \right) - \ln\left( \phantom{-}\del{x}_R + \frac{D_R}{2 \kappa} \right) \right)
    \, .
\end{align}
\end{subequations}
Since~$F_{L \to R}$ and~$F_{R \to L}$ depend on~$D_L$ and~$D_R$ respectively, and because~$F$ cannot be reduced further, we conclude that~$F_{L \to R} \neq F_{R \to L}$ and thus~$\sum{J} \neq 0$ as observed in \cref{sec:rw}.

\subsubsection{Sufficiently small time step}

For a sufficiently small time step~$\del{t}$, we can solve for the simplified expression of~$p_t$:
\begin{subequations}
\begin{align}
    F_{L \to R}
    &
    = \int_{-\del{x}_L}^{0} p_{t, L \to R}(x) \dl2{x}
    = \frac{2 \kappa}{D_L} \int_{-\del{x}_L}^{0} -x \dl2{x}
    = \frac{\kappa \del{x}_L^2}{D_L}
    = 2 \kappa \del{t}
    \, ,
    \\
    F_{R \to L}
    &
    = \int_{0}^{ \del{x}_R} p_{t, R \to L}(x) \dl2{x}
    = \frac{2 \kappa}{D_R} \int_{0}^{\del{x}_R} \phantom{-}x \dl2{x}
    = \frac{\kappa \del{x}_R^2}{D_R}
    = 2 \kappa \del{t}
    \, .
\end{align}
\end{subequations}
Hence,~$F_{L \to R} = F_{R \to L} = 2 \kappa \del{t}$ (we confirm this in simulations with small $\del{t}$) and the model will retain the steady-state solution. Recall that the simplification above requires that~$2 \kappa \del{s} \ll D$. This can be recast to
\begin{equation}
\label{eqn:dt_condition}
    1 \ll \frac{2 \kappa \sqrt{2 D \del{t}}}{D} = \sqrt{\frac{8 \kappa^2 \del{t}}{D}}
\end{equation}
and we now observe that the threshold at which a significant error in the fluxes occurs increases with~$\kappa$ and~$\del{t}$ and decreases with~$D$. This is in line with observations in~\cref{fig:rw_errorstudy}.

\subsection{Asymmetric interface reflection}
\label{sec:FluxAnalysis_interfacecondition}

As originally stated in~\cite{Szafer1995}, the interface reflection condition in \cref{eqn:interface_reflection} must be respected even as~$\kappa \to \infty$. This is satisfied intrinsically in \cref{eqn:pt_Maruyama_2sided}. Following from \cref{eqn:pt_Fieremans_2sided}, the ratio of probabilities
\begin{equation}
    \frac{p_{t, L \to R}}{p_{t, R \to L}}
    = \frac{2 \kappa \del{x}_{i, L}}{2 \kappa \del{x}_{i, R}} \frac{D_R + 2 \kappa \del{x}_{i, R}}{D_L + 2 \kappa \del{x}_{i, L}}
\end{equation}
%
approaches unity in the limit of~$\kappa \to \infty$ instead of $\sqrt{D_R/D_L}$. This might explain why the model appears to break down as the time step restriction is exceeded either by increasing~$\del{t}$ or~$\kappa$ as done in \cref{fig:rw_errorstudy}.

\section{Convergence}

\begin{figure}
    \centering
    \includegraphics{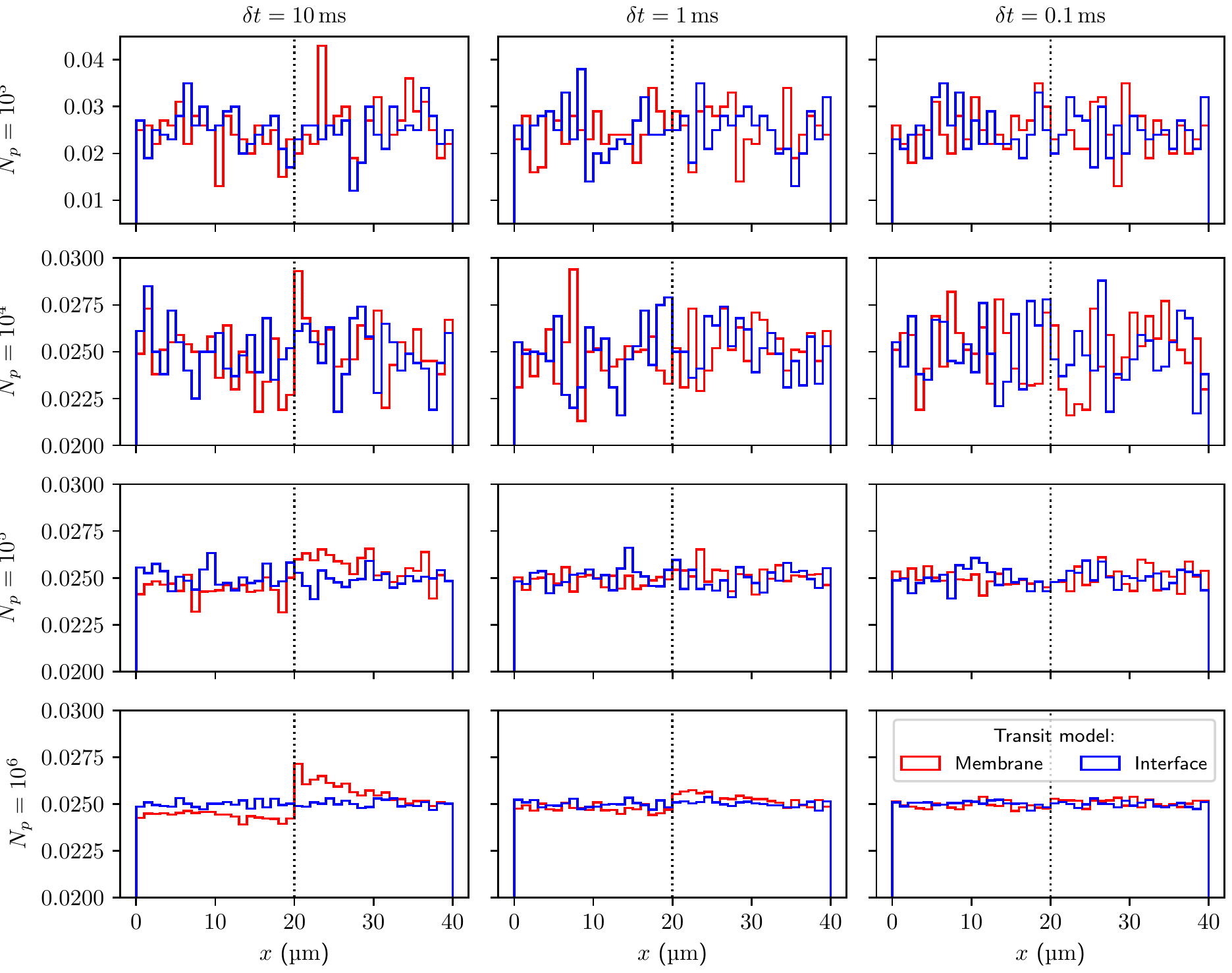}
    \caption{Convergence of random walk simulations of the steady-state in \cref{fig:rw_steadystate} for varying number of walkers~$N_p$ and time steps~$\del{t}$. Both transit models, the membrane model described in \cref{eqn:pt_Fieremans} and the interface model (which assumes $\kappa = \infty$) in \cref{eqn:pt_Maruyama}, are used. The total simulated time is~$t = \SI{100}{\unitT}$ and walkers are initially seeded uniformly in the domain ($D_L = \SI{2.5}{\unitD}$, $D_R = \SI{0.5}{\unitD}$, $\kappa = \SI{0.05}{\unitK}$).}
    \label{fig:rw_convergence}
\end{figure}

\Cref{fig:rw_convergence} shows the histograms of walker positions for both transit models considered in this work, \cref{eqn:pt_Fieremans,eqn:pt_Maruyama}. We consider three time steps and increase the number of walkers until the random fluctuations are of an acceptable level (compare \cref{sec:rw_steadystate}).

\end{appendices}
.

\end{document}